\documentclass[letterpaper,journal]{IEEEtran}
\usepackage{type1cm}
\usepackage{amsmath,amsfonts}
\usepackage{algorithmic}
\usepackage{array}
\usepackage[caption=false,font=footnotesize,labelfont=sf,textfont=sf]{subfig}
\usepackage{textcomp}
\usepackage{stfloats}
\usepackage{url}
\usepackage{verbatim}
\usepackage{graphicx}
\usepackage{array} 
\usepackage{amssymb}
\usepackage{siunitx}  
\usepackage{multirow}
\usepackage{booktabs} 
\usepackage{graphicx} 
\usepackage{soul,xcolor}
\usepackage{float} 
\usepackage{paralist}
\usepackage{marvosym}
\usepackage{fontawesome5}
\usepackage{tabularx}
\usepackage[switch]{lineno}
\usepackage{booktabs}
\usepackage{pifont}
\usepackage{makecell}
\def\BibTeX{{\rm B\kern-.05em{\sc i\kern-.025em b}\kern-.08em
    T\kern-.1667em\lower.7ex\hbox{E}\kern-.125emX}}
\usepackage{balance}

\usepackage[colorlinks=true, linkcolor=blue, citecolor=blue, urlcolor=blue, pdfborder={0 0 0}]{hyperref}

\begin{document}
\title{EvRWKV: A Continuous Interactive RWKV Framework for Effective Event-Guided Low-Light Image Enhancement}
\author{Wenjie Cai, Qingguo Meng\textsuperscript{\Letter}, Zhenyu Wang, Xingbo Dong, Zhe Jin
\thanks{
This work was supported in part by the National Natural Science Foundation of China under Grant 62376003 and Grant 62306003, in part by the Anhui Provincial Natural Science Foundation under Grant 2308085MF200, and in part by the Open Research Fund from Guangdong Laboratory of Artificial Intelligence and Digital Economy (SZ), under Grant No.GML-KF-24-29.
\textit{(Corresponding author: Qingguo Meng)}
}
\thanks{
Wenjie Cai is with Anhui Provincial International Joint Research center for Advanced technology in Medical imaging, School of Artificial Intelligence, Anhui University, Hefei 230601, China (e-mail: wa2214030@stu.ahu.edu.cn).

Qingguo Meng, and Zhe Jin are with State Key Laboratory of Opto-Electronic Information Acquisition and Protection Technology, the Anhui Provincial Key Laboratory of Secure Artificial Intelligence, Anhui Provincial International Joint Research Center for Advanced Technology in Medical Imaging, and the School of Artificial Intelligence, Anhui University, Hefei, China (e-mail: mqg1024@163.com; jinzhe@ahu.edu.cn).

Zhenyu Wang is with School of Artificial Intelligence, Anhui University, Hefei 230601, China (e-mail:wa2214026@stu.ahu.edu.cn).

Xingbo Dong is with State Key Laboratory of Opto-Electronic Information Acquisition and Protection Technology, Anhui Provincial International Joint Research Center for Advanced Technology in Medical Imaging, and the School of Artificial Intelligence, Anhui University, Hefei, China, and also with Guangdong Laboratory of Artificial Intelligence and Digital Economy (SZ), Shenzhen 518055, China (e-mail: xingbo.dong@ahu.edu.cn).

}}

\markboth{Journal of \LaTeX\ Class Files,~Vol.~18, No.~9, September~2020}%
{How to Use the IEEEtran \LaTeX \ Templates}

\maketitle

\begin{abstract}
Event cameras offer significant potential for Low-light Image Enhancement (LLIE), yet existing fusion approaches are constrained by a fundamental dilemma: early fusion struggles with modality heterogeneity, while late fusion severs crucial feature correlations. To address these limitations, we propose EvRWKV, a novel framework that enables continuous cross-modal interaction through dual-domain processing, which mainly includes a Cross-RWKV Module to capture fine-grained temporal and cross-modal dependencies, and an Event Image Spectral Fusion Enhancer (EISFE) module to perform joint adaptive frequency-domain denoising and spatial-domain alignment. This continuous interaction maintains feature consistency from low-level textures to high-level semantics. Extensive experiments on the real-world SDE and SDSD datasets demonstrate that EvRWKV significantly outperforms only image-based methods by 1.79 dB and 1.85 dB in PSNR, respectively. To further validate the practical utility of our method for downstream applications, we evaluated its impact on semantic segmentation. Experiments demonstrate that images enhanced by EvRWKV lead to a significant 35.44\% improvement in mIoU.
\end{abstract}


\begin{figure}[t] 
  \centering
  \includegraphics[width=\columnwidth]{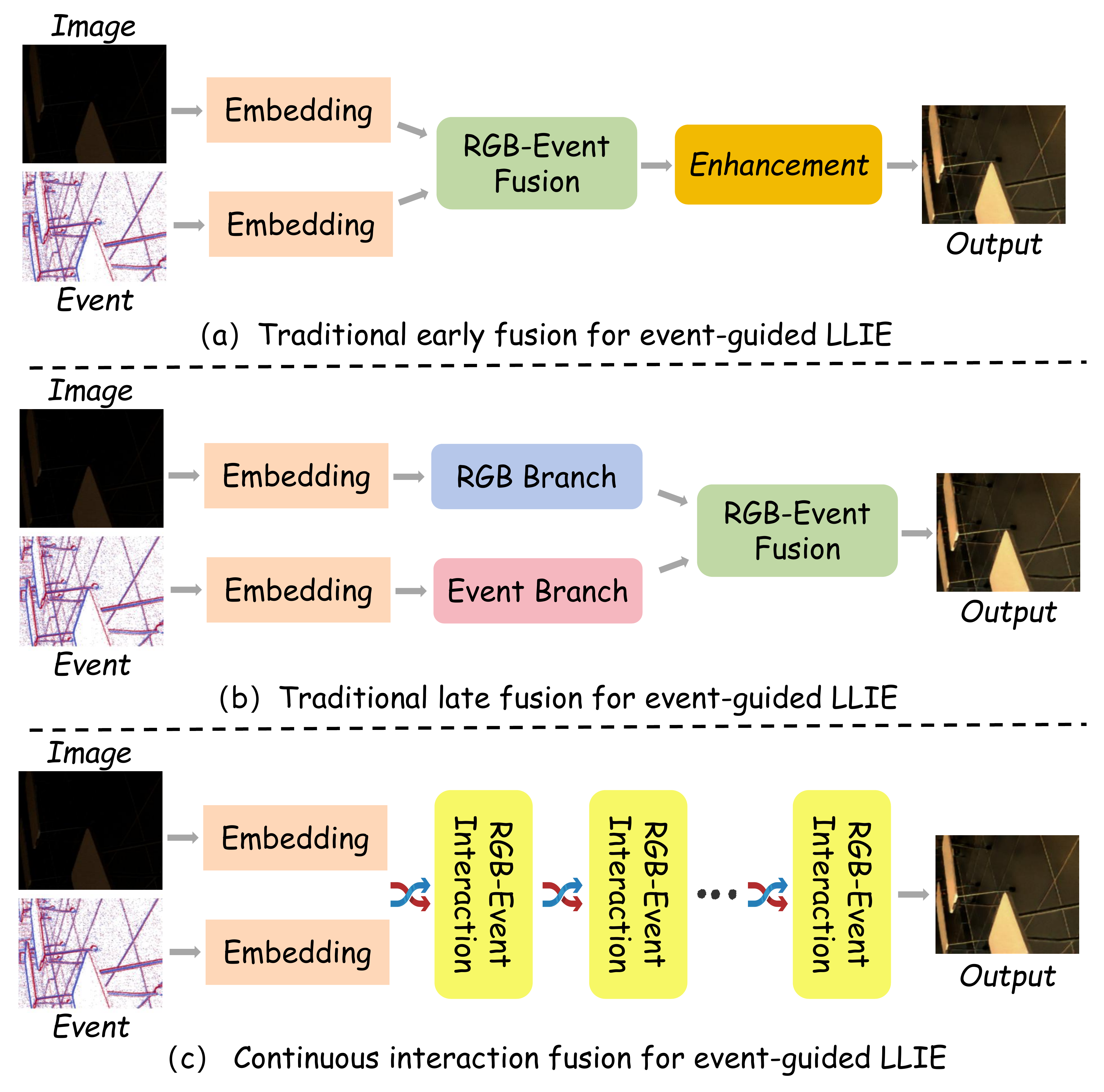} 
  \caption{Overall framework of event-guided LLIE approaches. (a) Early fusion: combining image and event data at the input. (b) Late fusion: processing image and event data separately and merging features. (c) Continuous interaction fusion: enabling ongoing interaction between image and event data.}
  \label{fig:1}
\end{figure}

\section{Introduction}
\label{label:intro}

\IEEEPARstart{C}{apturing} high-quality visual content under low-light conditions is a critical challenge in computer vision \cite{li2021low1}. Images acquired in such environments suffer from severe noise and underexposure \cite{li2021low1}, degrading downstream applications. While traditional frame-based low-light image enhancement (LLIE) methods have made progress \cite{lamba2021restoring,wang2021seeing,fu2016weighted, wang2023low}, they are often constrained by a fundamental trade-off. Attempts to brighten the image can severely amplify noise, while models trained on synthetic data often fail to generalize to the diverse and unpredictable nature of real-world scenes \cite{xu2022snr,cai2023retinexformer,wang2023ultra,10810469,li2021low}. 

Event cameras, using bio-inspired sensors \cite{lichtsteiner2008128}, are widely used in the field of computer vision due to their high dynamic range (HDR)  and microsecond level time resolution \cite{scheerlinck2019ced,zheng2023deep}, such as image restoration \cite{jiang2023event}, video super-resolution \cite{lu2023learning}, optical flow \cite{luo2023learning}, deblurring \cite{jiang2020learning,kim2022event,chen2025event}, etc., and emerge as a powerful solution for LLIE. By asynchronously capturing per-pixel brightness changes, events excel at preserving structural details even in extreme darkness and are less susceptible to motion blur. However, exploiting this synergy for LLIE is not straightforward. Methods \cite{stoffregen2020reducing,rebecq2019high} relying solely on event cameras are inherently limited by data sparsity, leading to poor spatial resolution and noise artifacts. Furthermore, fusion techniques combining events and images often use simplistic approaches \cite{jiang2023event,wang2020event,liu2023low,liang2024towards}, which fail to effectively capture the complex spatial-temporal interactions between them. As a result, these methods do not fully leverage the complementary strengths of event and image data, limiting their performance in LLIE tasks.

\begin{figure}[t] 
  \centering
  \includegraphics[width=\columnwidth]{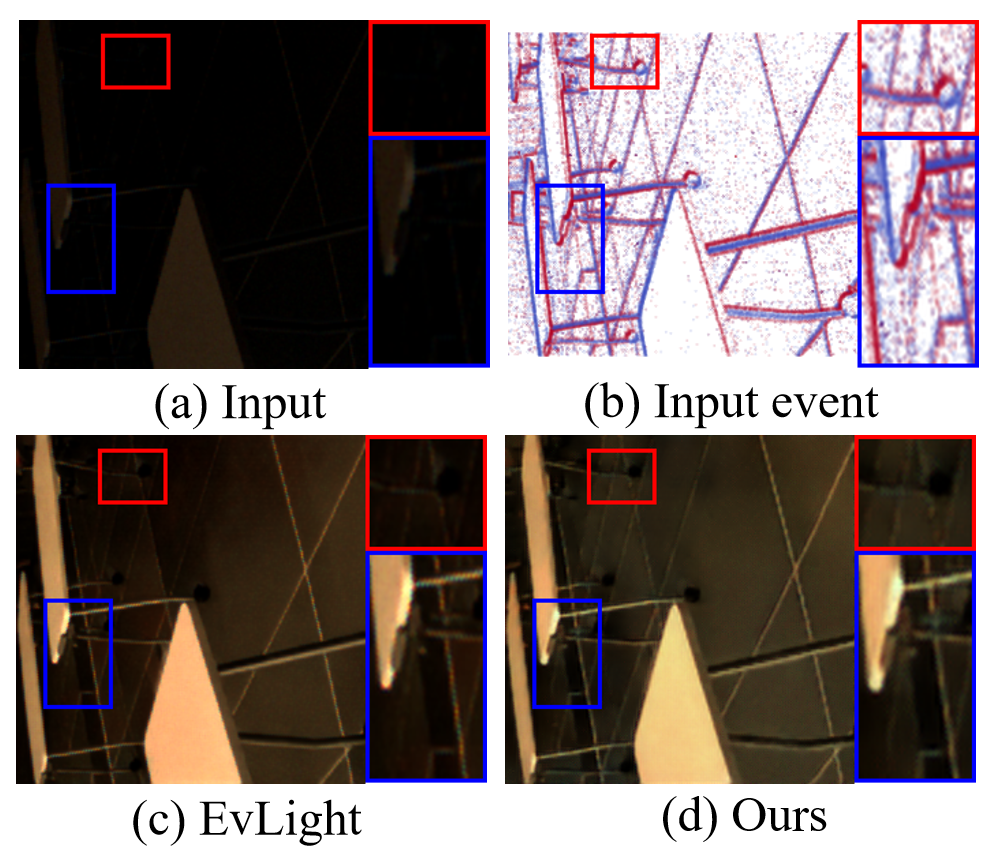} 
  \caption{A challenging example from our dataset containing an extremely low-light image (a) and sparse events (b). Compared with the result from the state-of-the-art event-guided method EvLight (c), our EvRWKV (d) not only recovers the wheel details in the dark areas (e.g., the wheel) but also preserves edge details (e.g., the white line on the floor).}
  \label{fig:first}
\end{figure}

The core challenge lies in navigating the fundamental dilemma posed by the two dominant fusion paradigms. As shown in Fig.~\ref{fig:1} (b), the prevailing methods often adopt a decoupled design \cite{liang2024towards,kim2024towards,sun2025low,liu2025bidirectional,jiang2023event}, processing the events and images through separate, parallel pathways. This architectural choice inherently inhibits continuous interaction, restricting feature exchange to a single terminal fusion step. Consequently, vital low-level complementary correlations are lost during independent hierarchical processing. These include transient event edges corresponding to faint image textures. By the time the features are combined, they are too abstract to benefit from this lost complementary detail, leading to underutilized complementarity and suboptimal enhancement, as shown in Fig.~\ref{fig:first} (c). Conversely, as shown in Fig.~\ref{fig:1} (a), strategies \cite{liang2023coherent,liu2023low} that merge modalities at the input level struggle with a fundamental mismatch in data structure. Events are temporally dense but spatially sparse \cite{lichtsteiner2008128}, whereas images offer dense spatial information, but are temporally discrete \cite{li2021low1}. Forcing these inherently complementary representations into a common format at the outset creates a representational bottleneck, obscuring their distinct characteristics and preventing the network from effectively learning their synergistic relationship. This approach may also amplify noise from both sources \cite{gallego2020event,li2021low1}, degrading the fused representation before feature extraction begins. Ultimately, the fundamental limitation across both paradigms is the absence of a continuous adaptive interaction between the modalities throughout the entire feature learning process.

To address this challenge, we introduce EvRWKV, a novel framework engineered for continuous cross-modal interaction through a dual-domain processing strategy. Instead of treating fusion as a single, discrete event, EvRWKV establishes a persistent dialogue between the image and event streams throughout the entire network hierarchy. At the core of our approach, the Cross-RWKV Module leverages the Receptance Weighted Key Value (RWKV) architecture \cite{peng2023rwkv} to enable fine-grained interaction between event and image features and ensuring consistent spatiotemporal alignment. Complementing this, the Event Image Spectral Fusion Enhancer (EISFE) module operates in a dual domain to jointly suppress noise, which is a critical issue in early fusion, while ensuring precise alignment of complementary structural details. By ensuring dynamic interaction across all stages of feature extraction, spanning from low-level textures to high-level semantics, EvRWKV achieves a seamless and holistic cross-modal fusion. By maintaining continuous feature alignment, this end-to-end collaboration prevents early information loss and late-stage disconnection, delivering a robust LLIE that effectively preserves fine details and suppresses noise and artifacts.

In summary, EvRWKV represents a significant advancement in LLIE by effectively harnessing the synergistic potential of event and image cameras through a unified, cross-modal, and cross-domain architecture. The contributions of this work are as follows:

\begin{itemize}
    \item We propose EvRWKV, a novel framework that establishes continuous cross-modal interaction through dual-domain processing for LLIE, effectively overcoming the limitations of both early and late fusion paradigms for robust low-light enhancement.

    \item We design Cross-RWKV, an RWKV-based backbone that maintains feature consistency from low-level to high-level to preserve and leverage cross-modal correlations.

    \item We introduce EISFE, a dual-domain module that performs joint noise mitigation and feature alignment, ensuring a robust and clean fusion process.

    \item We demonstrate that EvRWKV achieves state-of-the-art (SOTA) performance on LLIE across real-world datasets (SDE, SDSD, RELED), effectively suppressing noise and improving visual quality.

\end{itemize}


\section{Related Work}
\label{label:related}
\subsection{Frame-based LLIE}
LLIE aims to improve the visibility and quality of images captured in dimly lit environments. Traditional frame-based LLIE methods primarily operate on standard image frames. These methods can be broadly categorized into histogram equalization \cite{arici2009histogram,abdullah2007dynamic,celik2011contextual}, Gama correction \cite{huang2012efficient,rahman2016adaptive} and Retinex-based algorithms \cite{guo2016lime,xu2020star,qiang2025gwretinex,zhao2021retinexdip,10810469}. The first two methods directly enhance the intensity and contrast of low-light image, while the Retinex-based algorithm models the image as a combination of illumination and reflectivity. Although effective in improving visibility, these approaches often encounter limitations in handling complex lighting conditions, especially in scenes with varying illumination or dark regions. Moreover, these methods may amplify noise, particularly in low-light regions, leading to undesirable artifacts. 

Recent advancements have been propelled by deep learning, particularly with convolutional neural networks (CNNs) \cite{fu2016weighted,li2018structure,chen2018learning} and Transformer architectures \cite{xu2022snr,cai2023retinexformer,zhu2024ghost}, which have led to the development of more sophisticated frame-based LLIE methods. Chen et al. \cite{chen2018learning} established a low light image dataset and proposed a full convolution network for enhancement. Wang et al. \cite{wang2019underexposed} proposed to enhance underexposed photos by learning the illumination map. Wang et al. \cite{wang2021seeing} collect SDSD dataset by using mechatronic system and proposed a framework integrating progressive alignment and Retinex-based illumination prediction. Cai et al. \cite{cai2023retinexformer} propose a one-stage Transformer framework Retinexformer for LLIE, leveraging Retinex theory and illumination-guided attention to suppress noise. However, these frame-only methods fundamentally suffer from information loss in dark regions due to sensor noise floor limitations, leading to irreversible texture degradation and motion blur amplification.

\subsection{Event-guided LLIE}

The emergence of event cameras has significantly revolutionized low-light perception by offering a new paradigm in image sensing. Unlike traditional frame-based cameras that capture images at fixed intervals, event cameras detect changes in the scene asynchronously, producing a continuous stream of events that encode pixel-level intensity changes with high temporal resolution. Rebecq et al. \cite{rebecq2019high} propose a recurrent network to reconstruct high speed and high dynamic range videos from event camera data. Event-based approaches for LLIE like NER-Net \cite{liu2024seeing} leveraged event streams' temporal continuity to address non-uniform illumination via learnable timestamp calibration. Subsequent works explored hybrid event-frame fusion strategies. Jiang et al. \cite{jiang2023event} established joint feature learning through attention, while \cite{liang2024towards} introduced the SDE dataset with precise spatial-temporal alignment and proposed SNR-guided feature selection in EvLight framework. Recent advances in sensor 
fusion enhancement \cite{ragavendirane2025low} have demonstrated improved 
cross-modal integration through adaptive weighting mechanisms. Kim et al. \cite{kim2024towards} propose an event-guided end-to-end framework for joint low-light video enhancement and deblurring on the real-world RELED dataset, utilizing temporal alignment and cross-modal spectral filtering. Additionally, prior-guided approaches \cite{wang2025low} have shown promise in leveraging temporal consistency and historical information to stabilize enhancement under dynamic illumination. However, current fusion architectures are still limited by methods that rely on basic fusion strategies, which fail to effectively separate modality-specific noise patterns. Additionally, many of these approaches suffer from inflexible fusion operators that lack adaptive cross-modal interaction mechanisms. Furthermore, they often underutilize the event streams' intrinsic motion information, hindering comprehensive dynamic scene modeling.

\subsection{Receptance Weighted Key Value}
The Receptance Weighted Key Value (RWKV) architecture \cite{peng2023rwkv} emerges as a transformative sequential modeling paradigm, synergizing the parallel computation advantages of transformers with the memory efficiency of recurrent networks. Departing from conventional attention mechanisms constrained by quadratic complexity, RWKV introduces time-shifted receptance gates and exponentially decaying key projections to maintain hidden states \cite{peng2024eagle}, enabling linear computational scaling with sequence length. In computer vision, RWKV has demonstrated remarkable versatility across diverse tasks \cite{li2024survey}. For instance, Restore-RWKV \cite{yang2025restore} employs an omnidirectional token translation layer and recurrent WKV attention to restore medical images \cite{yang2025explicit} by capturing spatial dependencies, while PointRWKV \cite{he2025pointrwkv} leverages linear complexity and multi-head matrix states to enhance geometric feature extraction in 3D point cloud processing. Vision-RWKV \cite{duan2024vision} further extends its capabilities to high-resolution image analysis by reducing spatial aggregation complexity, outperforming traditional Vision Transformers (ViTs). In real-time video analytics, the TLS-RWKV framework \cite{zhu2024tls} achieves efficient temporal pattern recognition for action detection. 
StyleRWKV \cite{dai2024stylerwkv} utilizes RWKV models for efficient style transfer, enhancing global and local context while maintaining low memory usage and linear time complexity.

Despite these advances, the application of RWKV \cite{peng2023rwkv} to cross-modal fusion with asynchronous data streams remains largely unexplored. Effectively modeling such data requires addressing the dual challenges of temporal sparsity and asynchronicity \cite{gallego2020event}. While Transformers' all-to-all attention \cite{vaswani2017attention} is powerful, this un-differentiated global modeling can create spurious temporal dependencies by wrongly associating distant, unrelated events. On the other hand, conventional RNNs, with their data-dependent gating mechanisms, often exhibit training instability and struggle to capture long-range dependencies \cite{bengio1994learning} in highly sparse and noisy event data. The RWKV architecture is uniquely positioned to overcome these limitations. Its principled, non-learned time-decay mechanism provides a robust inductive bias, ensuring stable memory retention that aligns with the physical prior of a decaying event influence. Furthermore, its linear complexity and recurrent nature enable the native processing of long, continuous sequences, thereby preserving their asynchronous structure.  We propose a novel cross-modal synergy wherein the stable, temporally-aware backbone of RWKV allows events to resolve image ambiguities while images ground sparse event data, opening new pathways for applications in motion deblurring and LLIE.

\begin{figure*}[t]
  \centering
  \includegraphics[width=\textwidth]{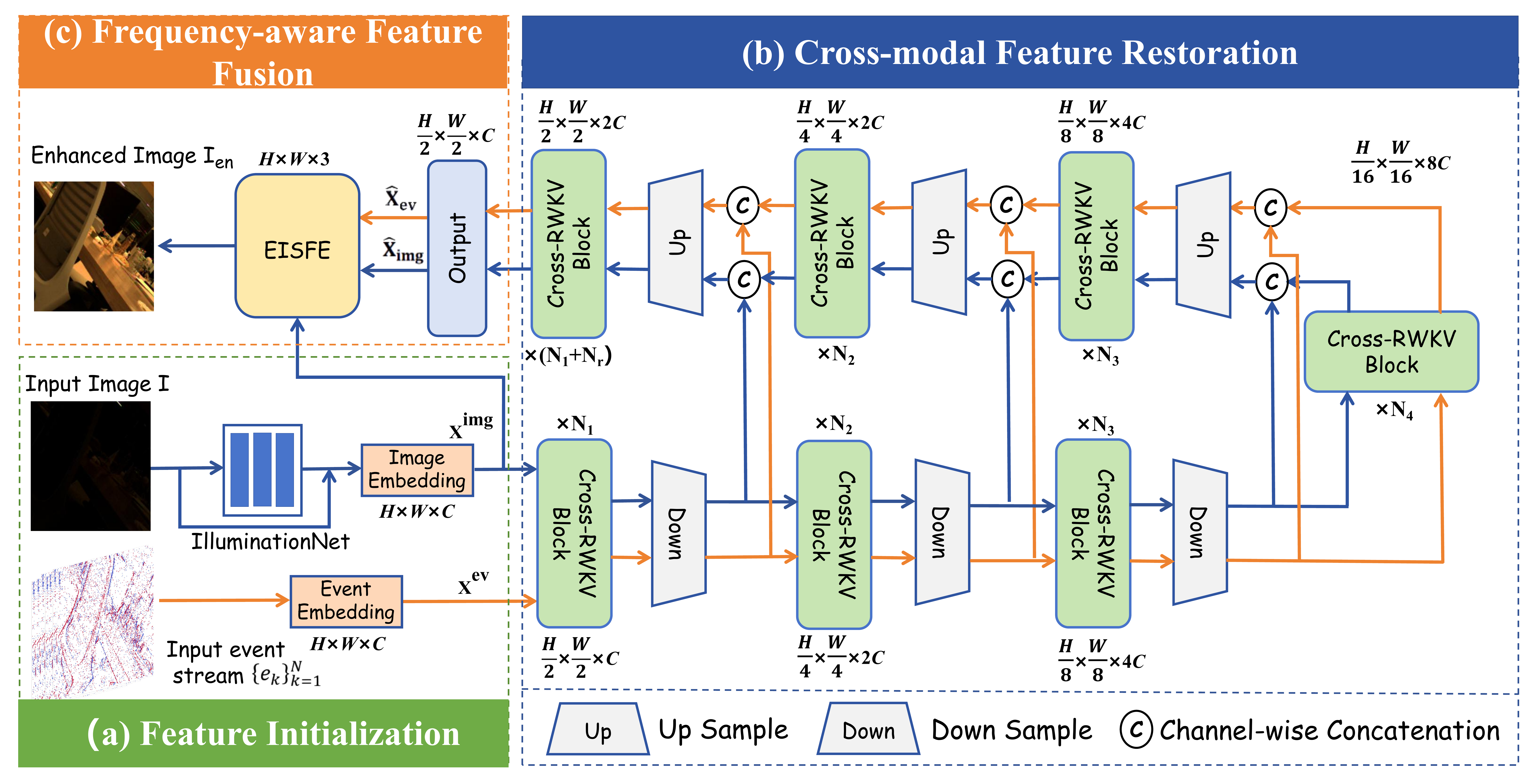} 
  \caption{Overall Architecture of the proposed EvRWKV. Our method consists of three parts: (a) Feature Initialization(Sec~\ref{label:a}), (b) Cross-modal Feature Restoration(Sec~\ref{label:b}), and (c) Frequency-aware Feature Fusion(Sec~\ref{label:c}). Specifically, Cross-modal Feature Restoration contains multiple Cross-RWKV blocks for feature alignment, and Frequency-aware Feature Fusion integrates image and event features for final output.}
  \label{Fig2}
\end{figure*}

\section{Methodology}
\label{label:method}
We propose \textit{EvRWKV}, a novel framework designed for continuous cross-modal interaction through dual-domain processin that leverages the complementary advantages of RGB data and event streams to achieve robust illumination enhancement and noise suppression. As shown in Fig. \ref{Fig2}, our framework comprises three key parts: 1) Feature Initialization, 2) Cross-modal Feature Restoration, 3) Frequency-aware Feature Fusion. The framework takes a low-light image \( I \) and paired event stream \( \{ e_k \}_{k=1}^{N} \) as inputs, producing an enhanced image \( I_{en} \). At first, a series of convolution layers extract features from low light images and event data, and then process them through the 4-level U-shaped encoder-decoder architecture composed of cross-RWKV composed of spatial and channel hybrid components. The output of the architecture passes the EISFE module is enhanced in frequency domain and spatial domain fusion processing to improve image quality.

\subsection{Feature Initialization}
\label{label:a}
During data loading, we first apply gamma correction for initial global brightness adjustment. Recent LLIE methods demonstrate that preliminary enhancement facilitates subsequent restoration. Following Retinex theory \cite{land1971lightness}, an image $I\in\mathbb{R}^{H\times W\times 3}$ can be decomposed into reflectance $R\in\mathbb{R}^{H\times W\times 3}$ and illumination $L\in\mathbb{R}^{H\times W}$ as $I = R \odot L$, where $\odot$ denotes element-wise multiplication. Beyond this global adjustment, following Retinexformer 
\cite{cai2023retinexformer}, we further enhance the 
gamma-corrected image $I$ through:
\begin{equation}
I_{lu} = I \odot \hat{L} + I,
\end{equation}
where $\hat{L}$ is the estimated illumination map, implementing adaptive enhancement while preserving original details.
The event stream $\{ e_k \}$ is converted into a voxel grid $E \in \mathbb{R}^{H \times W \times B}$ ($B = 32$) to preserve spatio-temporal information. The image $I_{lu}$ and event voxel $E$ are both downsampled to $\frac{H}{2}\times\frac{W}{2}$ via a two-layer 3$\times$3 convolution with strides 1 \& 2, extracting features $X^\text{img}, X^\text{ev} \in \mathbb{R}^{\frac{H}{2}\times\frac{W}{2}\times C}$ .

\subsection{Cross-modal Feature Restoration}
\label{label:b}
Existing methods for event-image fusion in low-light scenarios often fail to balance computational efficiency and robust cross-modal interaction, particularly in capturing high-frequency motion cues from events and preserving spatial details from images \cite{jiang2023event,liang2024towards}. Furthermore, they frequently treat event and image data as independent streams, failing to fully exploit the fine-grained spatiotemporal relationships between them. To address this, we propose the Cross-RWKV Module, which integrates Re-WKV \cite{yang2025restore} Attention for capturing global interactions and Cross-Modal Switch Omni-Shift (CS-Shift) for local feature enhancement and multi-modal information interaction. Specifically, multiple Cross-RWKV Modules form a 4-level encoder-decoder. The encoder progressively downsamples spatial dimensions and expands channels, while the decoder upsamples to restore resolution, enabling multi-scale cross-modal fusion. The outputs $O_{\text{img}},O_{\text{ev}}\in \mathbb{R}^{\frac{H}{2}\times\frac{W}{2}\times 2C}$ is split by a 3×3 convolution into $\hat{X}_{\text{img}}, \hat{X}_{\text{ev}}\in \mathbb{R}^{\frac{H}{2}\times\frac{W}{2}\times C}$. As shown in Fig.\ref{fig:2}, the module operates through spatial and channel mix, enabling efficient cross-modal fusion of event and image features. The details of the Cross-RWKV Block are as follows:
\subsubsection{Spatial Mix}
\begin{figure}[t] 
  \centering
  \includegraphics[width=0.75\columnwidth]{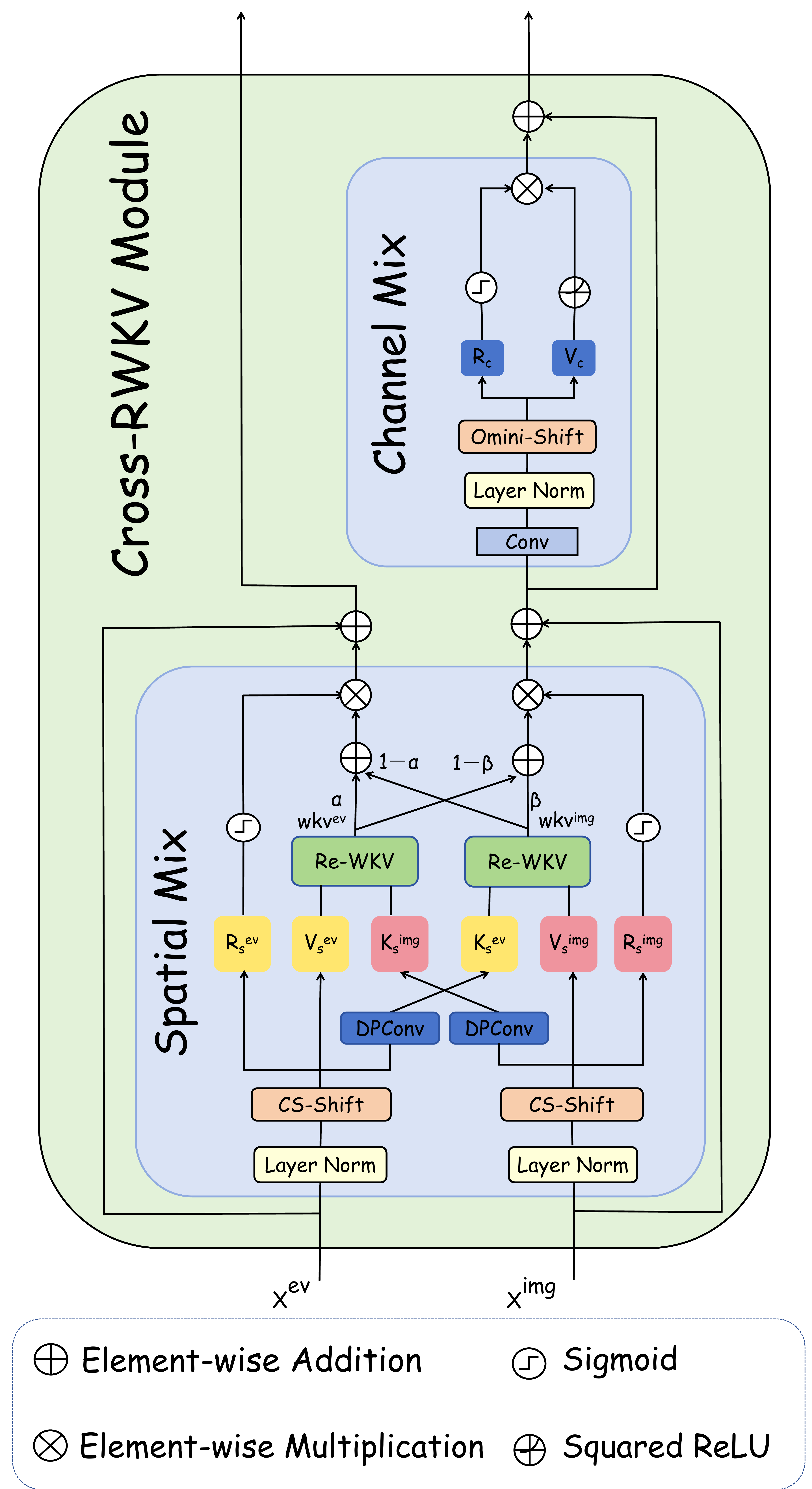} 
  \caption{Architecture of the Cross-RWKV Module, which includes Spatial Mix for spatial feature processing and Channel Mix for channel-wise interaction.
}
  \label{fig:2}
\end{figure}

The overall structure of Spatial Mix is illustrated in the Fig.\ref{fig:2}. This module aims to establish long-range dependencies across spatial dimensions while enabling effective interaction and information fusion between image and event modalities. Given the input feature sequences \( X^\text{img} \), \( X^\text{ev} \in \mathbb{R}^{T \times C} \), where \( T=\frac{H}{2} \times \frac{W}{2} \) represents the spatial resolution of tokens, the module processes image tokens and event tokens in parallel, first performing spatial-level feature fusion through Layer Normalization (LN) and the CS-shift mechanism.
\begin{equation}
X_{\text{s}}^{\text{img}}, X_{\text{s}}^{\text{ev}} = \text{CS-Shift}(LN_1(X^\text{img}), LN_2(X^\text{ev})),
\end{equation}

\noindent 
where the CS-Shift is formulated as:
\begin{equation}
  \begin{aligned}
  X_s^{\text{img}} &= \sum_{i=1}^{4} w_i^{\text{img}} \cdot \text{Conv}_{k\times k}(X_s^{\text{img}}), \\
  X_s^{\text{ev}} &= \sum_{i=1}^{4} w_i^{\text{ev}} \cdot \text{Conv}_{k\times k}(X_s^{\text{ev}}),
  \end{aligned}%
\end{equation}
where \( w_{i}^{\text{img}} \) and \( w_{i}^{\text{ev}} \) represent learnable parameters for scaling specific branches, and 
\( \text{Conv}_{k\times k}() \) denotes depthwise convolution with 
kernel sizes of \(1\times1\), \(3\times3\), and \(5\times5\), 
respectively. The scaling weights are initialized with random normal distribution 
and updated via backpropagation during training. Each CS-Shift module is learned 
independently per stage and per modality. During test, the four parallel 
branches are reparameterized into a single \(5\times5\) convolution for efficiency. 
This CS-shift mechanism captures both local and global features by fusing 
information under different receptive fields, resulting in accurate token shift 
outcomes.

Following the shift operation, instead of using standard linear projections, we adopt a depthwise separable convolution to compute keys:
\begin{equation}
    K_{s}^{\text{img}} = \text{DPConv}(X_{s}^{\text{img}}),  K_{s}^{\text{ev}} = \text{DPConv}(X_{s}^{\text{ev}}),
    \label{eq:key_img}
\end{equation}
 where  \( \text{DPConv}() \) consists of Depthwise Convolution and Pointwise Convolution, designed to preserve fine-grained spatial structures while efficiently expanding the receptive field.

Meanwhile, the value and receptance vectors for each modality are generated through linear projections:
\begin{equation}
    V_{s}^{\text{img}} = X_{s}^{\text{img}} W_{v}^{\text{img}}, \quad R_{s}^{\text{img}} = X_{s}^{\text{img}} W_{r}^{\text{img}},
\end{equation}
\begin{equation}
    V_{s}^{\text{ev}} = X_{s}^{\text{ev}} W_{v}^{\text{ev}}, \quad R_{s}^{\text{ev}} = X_{s}^{\text{ev}} W_{r}^{\text{ev}},
\end{equation}
where $W_{\text{img}}^{v}$, $W_{\text{img}}^{r}$, $W_{\text{ev}}^{v}$, and $W_{\text{ev}}^{r}$ are four fully connected layers.

\begin{figure}[t]
  \centering
  \includegraphics[width=0.75\columnwidth]{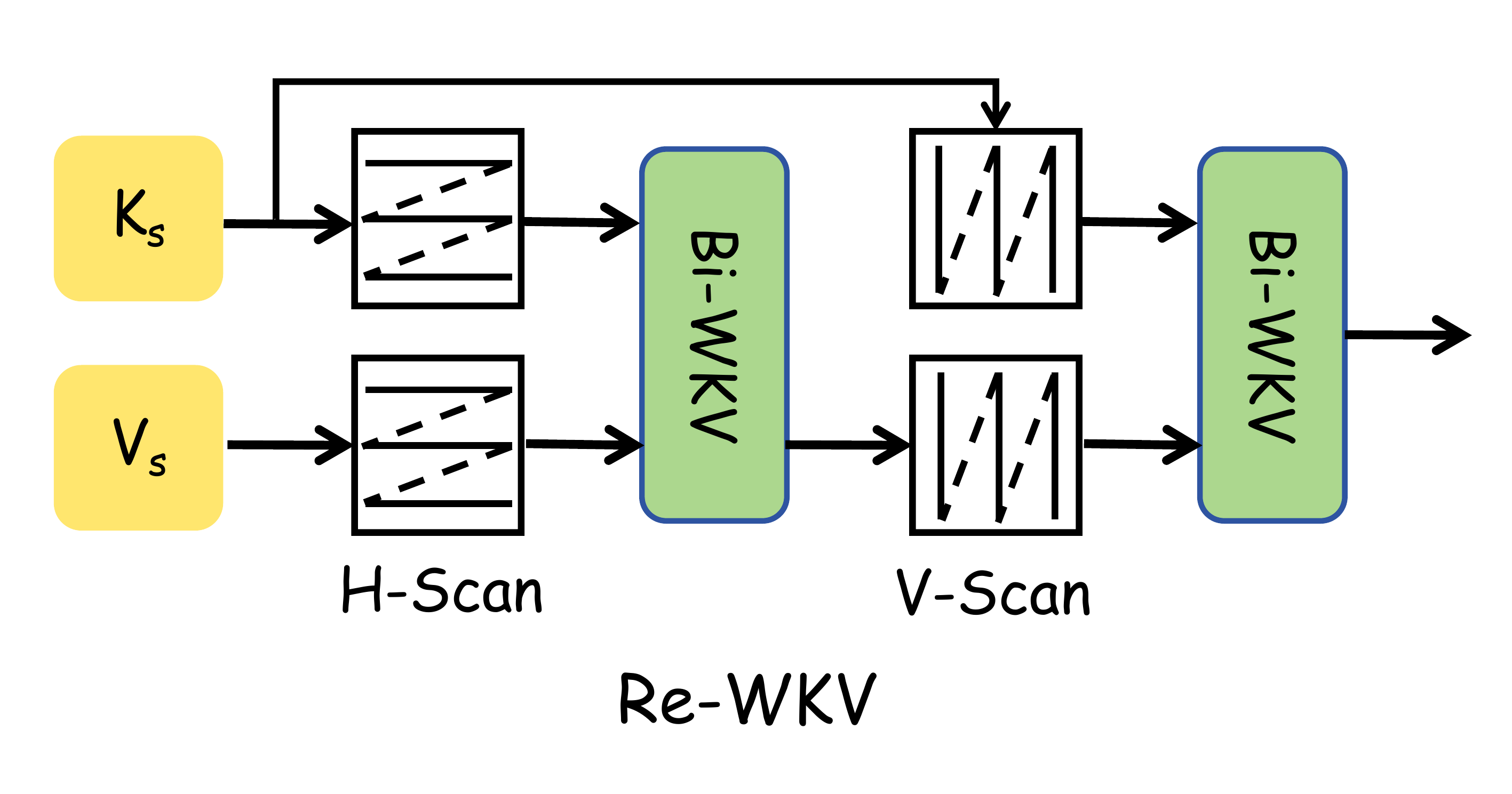}
  \caption{Illustration of Re-WKV mechanism. Bi-WKV attention is applied recurrently along alternating horizontal and vertical scan directions, with each iteration taking the previous attention result as input.}
  \label{fig:rewkv}
\end{figure}

Following the Restore-RWKV \cite{yang2025restore}, we employ the Re-WKV mechanism to compute the global attention result $\text{wkv} \in \mathbb{R}^{T \times C}$. At the same time, key representations are directly exchanged between image tokens and event tokens, allowing each modality to attend to the spatial structure of the other modality, thereby enabling joint spatial reasoning between images and events:
\begin{equation}
    \text{wkv}^{\text{img}} = \text{Re-WKV}(K_s^{\text{ev}}, V_s^{\text{img}}),
\end{equation}
\begin{equation}
    \text{wkv}^{\text{ev}} = \text{Re-WKV }(K_s^{\text{img}}, V_s^{\text{ev}}),
\end{equation}
where Re-WKV \cite{yang2025restore} is implemented by recurrently applying Bidirectional (Bi-WKV) attention along alternating scan directions. As shown in Fig.~\ref{fig:rewkv}, the recurrent process can be formulated as:
\begin{equation}
    \text{wkv}^{(j)} = \text{Bi-WKV}^{(j)}(\Delta_{\text{dir}}(K_s), \Delta_{\text{dir}}(\text{wkv}^{(j-1)})),
\end{equation}
where $\text{Bi-WKV}^{(j)}(\cdot)$ denotes the $j$-th Bi-WKV attention, $\Delta_{\text{dir}}(\cdot)$ represents the direction-changing operation that alternates between horizontal scan (H-Scan) and vertical scan (V-Scan), and $\text{wkv}^{(0)} = V_s$. The final Re-WKV output is $\text{wkv}^{(2)}$. Specifically, the Bi-WKV attention output for the $t$-th token is computed as:
\begin{equation}
\text{wkv}_t = \frac{\sum_{i=1,i\ne t}^{T} e^{-(|t-i|-1)/T \cdot (w + k_i)}v_i + e^{u + k_t}v_t}{\sum_{i=1,i\ne t}^{T} e^{-(|t-i|-1)/T \cdot (w + k_i)} + e^{u + k_t}}.
\end{equation}

Here, $T$ denotes the total number of tokens, $k_i$ and $v_i$ are the key and value vectors at position $i$, and $w \in \mathbb{R}^C$, $u \in \mathbb{R}^C$ are learnable parameters that control the relative positional bias and provide additional weight to the current token, respectively. By applying Bi-WKV iteratively across multiple directions, Re-WKV \cite{yang2025restore} effectively models long-range dependencies in 2D spatial structures, enabling enhanced global contextual reasoning across modalities.

To further promote cross-modal feature alignment, we introduce learnable fusion gates $\alpha_{\text{img}}, \alpha_{\text{evt}} \in \mathbb{R}^C$, applied to modulate the contribution from each modality during the final representation fusion:
\begin{equation}
    X_1 = \sigma(\alpha_{\text{img}}) \cdot \text{wkv}^{\text{img}} + \left(1 - \sigma(\alpha_{\text{img}})\right) \cdot X^{\text{ev}},
\end{equation}
\begin{equation}
    X_2 = \sigma(\alpha_{\text{evt}}) \cdot \text{wkv}^{\text{ev}} + \left(1 - \sigma(\alpha_{\text{evt}})\right) \cdot X^{\text{img}}.
\end{equation}

These fused representations are then gated by the learned receptance signals to produce the final outputs:
\begin{equation}
    O_{\text{img}} = \left( \sigma(R_{\text{img}}) \odot X_1 \right) W_o^{\text{img}},
\end{equation}
\begin{equation}
    O_{\text{evt}} = \left( \sigma(R_{\text{evt}}) \odot X_2 \right) W_o^{\text{evt}},
\end{equation}
where $W_o^{\text{img}}$ and $W_o^{\text{evt}}$ are linear output projection matrices.

This design enables explicit cross-modal interaction through key exchange, spatially-aware recurrent updates via Re-WKV, and adaptive fusion through gating, yielding a robust mechanism for unified spatial modeling over both image and event streams.

\begin{figure}[t] 
  \centering
  \includegraphics[width=\columnwidth]{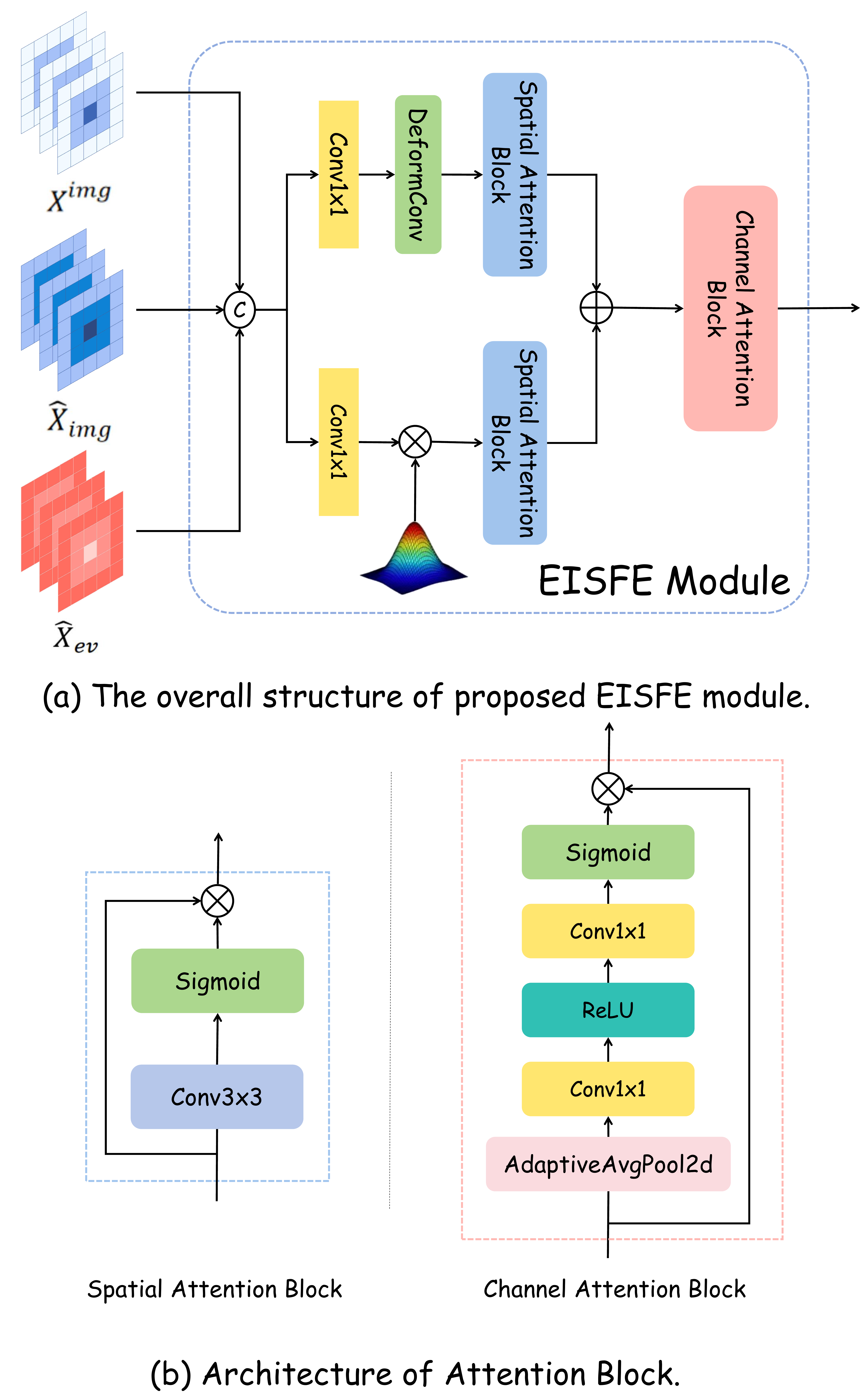} 
  \caption{Overview of the proposed EISFE module. (a) The overall structure includes Channel and Spatial Attention Blocks combined with an Adaptive Gaussian Filter to fuse inputs from image and event features. (b) Details of the Channel Attention Block and Spatial Attention Block architectures.}
  \label{fig:3}
\end{figure}

\subsubsection{Channel Mix}

The Channel Mix module is designed to model cross-channel interactions within each spatial token while preserving the spatial structure of visual inputs. Unlike the Spatial Mix module that jointly processes both image and event modalities, Channel Mix focuses solely on the image stream. This design choice stems from the observation that channel-wise semantic structures are generally more informative and stable in image data than in sparse and noisy event representations.

Given the image features $X_{\text{img}} \in \mathbb{R}^{T \times C}$, where $T = H \times W$ denotes the number of spatial tokens and $C$ is the channel dimension, the input first undergoes an Omni-Shift operation to enhance local contextual information:
\begin{equation}
X_{\text{c}}^{\text{img}} = \text{OmniShift}(\text{Conv}(X_{\text{img}})).
\end{equation}

The shifted feature is then passed through a gated channel-wise mixing pipeline. A key projection is first applied:
\begin{equation}
K_{\text{c}} = \text{squared ReLU}(X_{\text{c}}^{\text{img}} W_k),
\end{equation}
where $W_k \in \mathbb{R}^{C \times C_h}$ is the key projection matrix and the squared ReLU non-linearity emphasizes strong activations while suppressing weak ones. This is followed by a value projection to obtain the transformed feature:
\begin{equation}
V_{\text{c}} = K_{\text{c}} W_v, \quad W_v \in \mathbb{R}^{C_h \times C}.
\end{equation}

A separate gating signal is generated through a linear transformation and a sigmoid activation:
\begin{equation}
R_{\text{c}} = \sigma(X_{\text{c}}^{\text{img}} W_r), \quad W_r \in \mathbb{R}^{C \times C}.
\end{equation}

The final output is obtained via element-wise modulation:
\begin{equation}
O_{\text{c}} = R_{\text{c}} \odot V_{\text{c}}.
\end{equation}

By restricting Channel Mix to the image modality only, we ensure more stable and informative feature transformations. Event streams, while rich in motion cues, often suffer from high sparsity and noise in the channel dimension, making them less suitable for dense feedforward modeling. Therefore, reserving event data for interaction in the spatial domain while using only image data for channel-wise transformations leads to a better division of modeling responsibilities, improving overall robustness and representation quality.

\subsection{Frequency-aware Feature Fusion}
\label{label:c}

To effectively integrate complementary cues from image and event modalities, we propose the EISFE module, a Frequency-aware Image-Event Spatial-Frequency Enhancement module. This dual-domain fusion mechanism jointly exploits frequency-domain smoothness and spatial-domain adaptivity to enhance reconstruction quality, particularly in challenging scenarios such as motion blur, fast motion, or low-light degradation. 

Given the restored image feature $\hat{X}_{\text{img}}$, event feature $\hat{X}_{\text{ev}}$, and raw image input $X^{\text{img}}$, we first concatenate them along the channel dimension to form a unified representation:
\begin{equation}
X_{\text{fused}} = \text{Concat}(\hat{X}_{\text{img}}, \hat{X}_{\text{ev}}, X^{\text{img}}).
\end{equation}

We then apply two separate $1 \times 1$ convolutions to decompose $X_{\text{fused}}$ into two parallel branches: a frequency branch $X_{\text{freq}}$ and a spatial branch $X_{\text{spat}}$.

In the frequency branch, we introduce a channel-wise Adaptive Gaussian Filtering mechanism to selectively suppress noise while preserving structural information. Unlike traditional fixed-filter approaches, our method learns a distinct Gaussian standard deviation $\sigma_c$ for each channel, enabling content-adaptive filtering across different frequency sensitivities. Given the frequency-branch input $X_{\text{freq}} \in \mathbb{R}^{C \times H \times W}$, we generate a 2D Gaussian kernel $G_{\sigma_c} \in \mathbb{R}^{K \times K}$ for each channel using its learned $\sigma_c \in [\sigma_{\min}, \sigma_{\max}]$, and define it as:

\begin{equation}
G_{\sigma_c}(x, y) = \frac{1}{2\pi\sigma_c^2} \exp\left( -\frac{x^2 + y^2}{2\sigma_c^2} \right),
\end{equation}
where $(x, y)$ are spatial coordinates.

To efficiently apply filtering, we perform the convolution in the frequency domain via Fast Fourier Transform (FFT). Specifically, for each channel $c$, the filtered output is computed as:
\begin{equation}
\hat{X}_{\text{freq}, c} = \mathcal{F}^{-1} \left( \mathcal{F}(X_{\text{freq}, c}) \cdot \mathcal{F}(G_{\sigma_c}) \right),
\end{equation}
where $\mathcal{F}(\cdot)$ and $\mathcal{F}^{-1}(\cdot)$ denote 2D FFT and inverse FFT, respectively. This formulation allows the model to perform global frequency-aware filtering with high efficiency and adaptivity.

By learning $\sigma_c$ per channel, this module dynamically adjusts filtering strength: it smooths flat areas while preserving edge structures and fine details, making it well-suited for denoising and contrast enhancement in degraded conditions.

In parallel, the spatial branch processes $X_{\text{spat}}$ using a deformable convolution. This allows flexible receptive fields that adapt to local content, enabling the model to capture fine-grained spatial variations such as motion boundaries or geometric distortions. The output is denoted as:
\begin{equation}
\hat{X}_{\text{spat}} = \text{DeformConv}(X_{\text{spat}}),
\end{equation}

which retains high-frequency spatial patterns that complement the globally smoothed features from the frequency branch.

To integrate the outputs of the frequency and spatial branches, we design a hierarchical attention fusion strategy that sequentially applies spatial and channel attention mechanisms. This allows the model to selectively emphasize salient patterns while suppressing irrelevant noise, thereby enhancing cross-modal feature integration. As illustrated in Fig.~\ref{fig:3}, the fusion begins by computing spatial attention maps independently for both branches:

\begin{equation}
X_{\text{attn-spat}} = A_{\text{freq}} \odot \hat{X}_{\text{freq}} + A_{\text{spat}} \odot \hat{X}_{\text{spat}},
\end{equation}
where $A_{\text{freq}},A_{\text{spat}}$ denotes spatial attention maps independently
for both branches, and $\odot$ represents element-wise multiplication. This formulation adaptively combines spatial details and frequency-aware structures based on per-pixel relevance.

Following the spatial attention stage, we further refine the fused features through channel-wise attention to emphasize globally informative dimensions. As shown in Fig.~\ref{fig:3}, we first aggregate global context via adaptive global average pooling:
\begin{equation}
X_{\text{attn}} = X_{\text{attn-spat}} \odot A_{\text{chan}}.
\end{equation}

This two-stage attention mechanism allows the EISFE module to dynamically adapt to both local and global feature importance. The spatial attention modulates each spatial location based on semantic content, while the channel attention globally reweighs feature maps to highlight informative channels. Together, these mechanisms enable robust and adaptive fusion of image-event representations under various degradation conditions.

Finally, the fused feature map \(X_{\text{attn}}\in\mathbb{R}^{C\times \frac{H}{2}\times \frac{W}{2}}\)
is first linearly projected with a \(1\times 1\) convolution to refine channel
interactions, then up–sampled to the original spatial resolution by a learnable
\(4\times 4\) transposed convolution, and finally remapped
to the RGB space through another \(1\times 1\) convolution.

\subsection{Loss Function}  \label{sec:loss}

The loss function $\mathcal{L}$ used for training our EvRWKV model is composed of four complementary terms, formulated as:
\begin{equation}
  \mathcal{L}_{\text{total}} = \lambda_r \mathcal{L}_r + \lambda_p \mathcal{L}_p + \lambda_s \mathcal{L}_s + \lambda_m \mathcal{L}_m,
\end{equation}
where $\lambda_r = 1$, $\lambda_p = 0.8$, $\lambda_s = 1$, and $\lambda_m = 0.5$ are balancing weights. The reconstruction term combines pixel-wise fidelity and perceptual similarity:
\begin{equation}
\begin{aligned}
  \mathcal{L}_r =  \sqrt{(I_{\text{en}} - I_{\text{gt}})_i^2 + \epsilon^2}, \\
  \mathcal{L}_p = \|\Phi(I_{\text{en}}) - \Phi(I_{\text{gt}})\|_1,
  \end{aligned}
  \label{eq:reconstruction}
\end{equation}
where $\epsilon = 10^{-4}$ and $\Phi$ denoting AlexNet-based feature extraction. 

To ensure enhanced images maintain natural structural characteristics and visual quality, we employ both the Structural Similarity (SSIM) loss and its multi-scale extension (MS-SSIM):

\begin{equation}
  \begin{aligned}
 \mathcal{L}_s = 1 - \text{SSIM}(I_{\text{en}}, I_{\text{gt}}), \\ 
  \mathcal{L}_m = 1 - \text{MS-SSIM}(I_{\text{en}}, I_{\text{gt}}). 
  \end{aligned}
  \label{eq:structural}
\end{equation}

\begin{table*}[tbp]
\renewcommand{\arraystretch}{1.3}
\centering
\fontsize{7.5}{10}\selectfont
\caption{Quantitative comparison of different methods on the SDE dataset. 
The best and second-best results are highlighted in bold and underlined, respectively.}
\resizebox{\textwidth}{!}{  
\begin{tabular}{
>{\centering\arraybackslash}p{3.13cm}
>{\centering\arraybackslash}p{1.2cm}
*{2}{c}
*{4}{c}
*{4}{c}}
\Xhline{1.5pt}
\multirow{2}{*}{\textbf{Method}} & 
\multirow{2}{*}{\textbf{Backbone}} &
\multicolumn{2}{c}{\textbf{Inputs}}&
\multicolumn{4}{c}{\textbf{SDE-in}}& 
\multicolumn{4}{c}{\textbf{SDE-out}}\\
\cline{3-12}
& & \textbf{Image} & \textbf{Event}
& \textbf{PSNR$\uparrow$} & \textbf{SSIM$\uparrow$} & \textbf{LPIPS$\downarrow$} & \textbf{NIQE$\downarrow$}
& \textbf{PSNR$\uparrow$} & \textbf{SSIM$\uparrow$} & \textbf{LPIPS$\downarrow$} & \textbf{NIQE$\downarrow$} \\
\hline
SNR-Net \cite{xu2022snr} (CVPR'22) & Transformer & \ding{51} & \ding{55} 
& 20.05 & 0.630 & 0.244 & 14.86 
& 22.18 & 0.661 & 0.184 & 12.16 \\

Uformer \cite{wang2022uformer} (CVPR'22) & Transformer & \ding{51} & \ding{55} 
& 21.09 & 0.752 & 0.109 & \underline{9.670} 
& 22.32 & \underline{0.747} & 0.097 & \underline{8.969} \\

LLFlow \cite{wu2023learning} (CVPR'23) & CNN & \ding{51} & \ding{55} 
& 20.92 & 0.661 & 0.225 & 11.76 
& 21.68 & 0.647 & 0.236 & 14.56 \\

Retinexformer \cite{cai2023retinexformer} (ICCV'23) & Transformer & \ding{51} & \ding{55} 
& 21.30 & 0.692 & 0.124 & 11.24 
& \underline{22.92} & 0.683 & 0.146 & 11.35 \\

RetinexMac \cite{10810469} (TCSVT'25) & Transformer & \ding{51} & \ding{55} 
& 20.61 & 0.650 & 0.149 & 12.54 
& 21.68 & 0.683 & 0.170 & 11.12 \\
\hline

eSL-Net \cite{wang2020event} (ECCV'20) & CNN & \ding{51} & \ding{51} 
& 21.42 & 0.725 & 0.130 & 10.62 
& 21.39 & 0.681 & 0.282 & 15.81 \\

ELIE \cite{jiang2023event} (TMM'23) & Transformer & \ding{51} & \ding{51} 
& 19.98 & 0.617 & 0.217 & 14.75 
& 20.69 & 0.653 & 0.313 & 15.98 \\

LLVE-SEG \cite{liu2023low} (AAAI'23) & Transformer & \ding{51} & \ding{51} 
& 21.79 & 0.705 & 0.114 & 10.54 
& 22.35 & 0.690 & 0.154 & 12.39 \\

ELEDNet \cite{kim2024towards} (ECCV'24) & Transformer & \ding{51} & \ding{51} 
& 21.46 & 0.713 & 0.116 & 13.60 
& 22.91 & 0.726 & 0.129 & 12.77 \\

EvLight \cite{liang2024towards} (CVPR'24) & Transformer & \ding{51} & \ding{51} 
& \underline{22.21} & \underline{0.758} & \underline{0.101} & 11.13 
& 22.13 & 0.725 & \textbf{0.090} & 11.48 \\
\hline

\textbf{Ours} & RWKV & \ding{51} & \ding{51} 
& \textbf{23.09} & \textbf{0.770} & \textbf{0.094} & \textbf{8.534} 
& \textbf{23.43} & \textbf{0.768} & \underline{0.091} & \textbf{8.926} \\
\Xhline{1.5pt}
\end{tabular}}
\label{tab:1}
\end{table*}

\begin{table*}[tbp]
\renewcommand{\arraystretch}{1.3}
\centering
\fontsize{7.5}{10}\selectfont
\caption{Quantitative comparison of different methods on the SDSD and RELED datasets. 
The best and second-best results are highlighted in bold and underlined, respectively.}
\resizebox{\textwidth}{!}{
\begin{tabular}{
>{\centering\arraybackslash}p{2cm}
*{4}{c}
*{4}{c}
*{4}{c}}
\Xhline{1.5pt}
\multirow{2}{*}{\textbf{Method}} & 
\multicolumn{4}{c}{\textbf{SDSD-in}} & 
\multicolumn{4}{c}{\textbf{SDSD-out}} &
\multicolumn{4}{c}{\textbf{RELED}} \\
\cline{2-13}
& \textbf{PSNR$\uparrow$} & \textbf{SSIM$\uparrow$} & \textbf{LPIPS$\downarrow$} & \textbf{NIQE$\downarrow$}
& \textbf{PSNR$\uparrow$} & \textbf{SSIM$\uparrow$} & \textbf{LPIPS$\downarrow$} & \textbf{NIQE$\downarrow$}
& \textbf{PSNR$\uparrow$} & \textbf{SSIM$\uparrow$} & \textbf{LPIPS$\downarrow$} & \textbf{NIQE$\downarrow$} \\
\hline
SNR-Net \cite{xu2022snr}
& 24.74 & 0.830 & 0.097 & 7.963 
& 24.82 & 0.740 & 0.107 & 4.638
& 26.47 & 0.851 & 0.192 & 14.76 \\

Uformer \cite{wang2022uformer} 
& 26.60 & 0.881 & 0.068 & 6.607 
& 24.08 & 0.818 & 0.137 & 4.773
& 28.86 & 0.832 & 0.212 & 14.40 \\

LLFlow \cite{wu2023learning} 
& 23.39 & 0.818 & 0.104 & 9.586 
& 20.39 & 0.634 & 0.255 & 7.468
& 28.62 & 0.862 & 0.154 & 13.07 \\

Retinexformer \cite{cai2023retinexformer} 
& 25.90 & 0.852 & 0.086 & 6.954 
& \underline{26.08} & 0.815 & 0.061 & \underline{4.322}
& 26.66 & 0.865 & 0.168 & 13.54 \\

RetinexMac \cite{10810469} 
& 27.61 & 0.900 & 0.046 & \underline{6.550} 
& 24.75 & 0.783 & 0.103 & 5.372
& 28.55 & 0.851 & 0.192 & 12.26 \\
\hline

eSL-Net \cite{wang2020event} 
& 23.68 & 0.821 & 0.067 & 7.299 
& 24.95 & 0.805 & 0.063 & 4.821
& 25.76 & 0.796 & 0.312 & 16.11 \\

ELIE \cite{jiang2023event} 
& 27.46 & 0.879 & 0.074 & 6.736 
& 23.29 & 0.742 & 0.116 & 5.568
& 26.62 & 0.862 & 0.267 & 14.14 \\

LLVE-SEG \cite{liu2023low} 
& 27.58 & 0.888 & 0.053 & 6.837 
& 23.51 & 0.726 & 0.110 & 5.036
& 29.19 & 0.875 & 0.098 & 11.94 \\

ELEDNet \cite{kim2024towards}
& 28.50 & 0.910 & 0.040 & 7.439 
& 25.17 & 0.817 & 0.081 & 4.895
& \underline{31.96} & \textbf{0.910} & 0.106 & 10.82 \\

EvLight \cite{liang2024towards}
& \underline{28.52} & \underline{0.912} & \underline{0.039} & 6.764 
& 25.08 & \underline{0.828} & \underline{0.060} & 4.455
& 31.29 & \underline{0.885} & \underline{0.067} & \underline{10.63} \\
\hline

\textbf{Ours} 
& \textbf{28.96} & \textbf{0.920} & \textbf{0.032} & \textbf{6.451} 
& \textbf{27.93} & \textbf{0.839} & \textbf{0.058} & \textbf{4.110}
& \textbf{32.18} & \textbf{0.910} & \textbf{0.044} & \textbf{9.207} \\
\Xhline{1.5pt}
\end{tabular}}
\label{tab:2}
\end{table*}

\section{EXPERIMENTS}
\label{label:experiments}

\subsection{Experimental Settings}

\subsubsection{Datasets and Evaluation Metrics}

SDE \cite{liang2024towards}, SDSD \cite{wang2021seeing}, and RELED \cite{kim2024towards} datasets are selected to test the enhancement performance of our EvRWKV framework.

\textbf{SDE} dataset contains 91 image–event paired sequences (43 indoor and 48 outdoor) captured with a DAVIS346 event camera at a resolution of \(346 \times 260\); it is primarily used to evaluate the capability of different methods to leverage event cues for LLIE under diverse illumination conditions. 

\textbf{SDSD} dataset provides 150 low-light/normal-light paired video sequences with an original resolution of \(1920 \times 1080\); following the dynamic-subset split (125 sequences for training and 25 for testing), all videos are down-sampled to \(346 \times 260\), and their event streams are synthesized with the v2e \cite{hu2021v2e} simulator. This dataset evaluates the robustness of enhancement methods in dynamic low-light scenes. 

\textbf{RELED} dataset comprises 42 scenes (29 for training and 13 for testing), each offering a low-light blurred input directly paired with a sharp, normal-light reference, making it suitable for benchmarking algorithms that jointly address illumination enhancement and deblurring. 

\begin{figure*}[t]
  \centering
  \includegraphics[width=\textwidth]{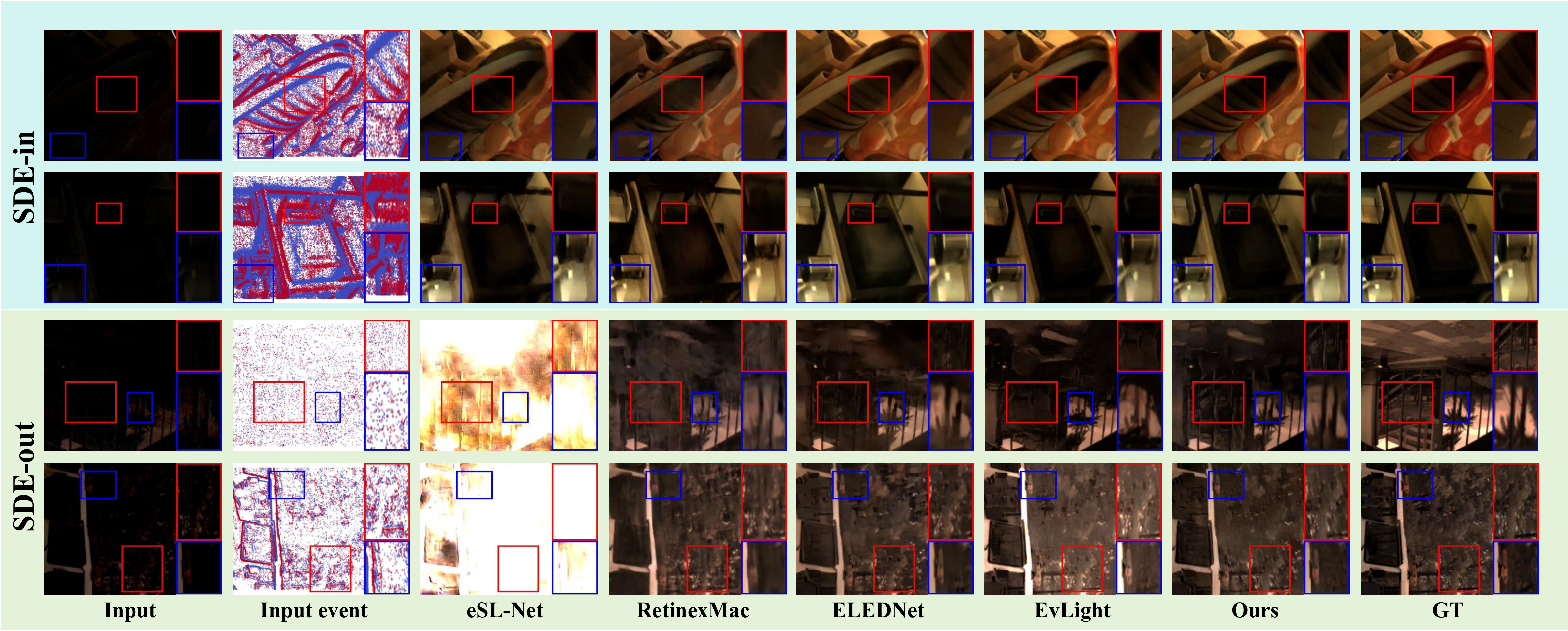} 
  \caption{Visual comparisons  with state-of-the-art methods on SDE-in and SDE-out dataset \cite{liang2024towards}.}
  \label{fig:sde}
\end{figure*}

\begin{figure*}[t]
  \centering
  \includegraphics[width=\textwidth]{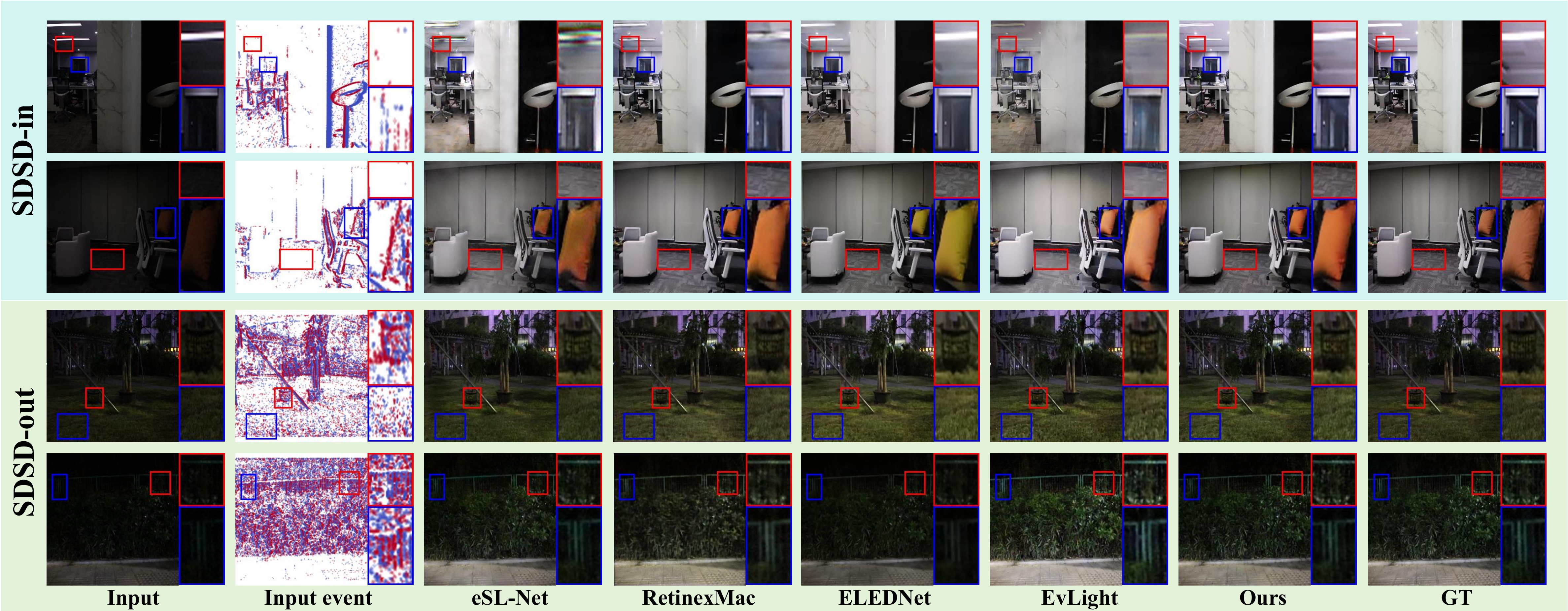} 
  \caption{Visual comparisons  with state-of-the-art methods on SDSD-in and SDSD-out dataset \cite{wang2021seeing}.}
  \label{fig:sdsd}
\end{figure*}

Together, these three benchmarks provide a comprehensive testbed for assessing the multi-modal LLIE capability of our EvRWKV method. We use the peak-signal-to-noise ratio (PSNR) \cite{hore2010image}, structural similarity index (SSIM) \cite{wang2004image}, learned perceptual image patch similarity (LPIPS) \cite{zhang2018unreasonable}, and natural image quality evaluator (NIQE) \cite{mittal2012making} for quantitative evaluation, where higher PSNR and SSIM values and lower LPIPS and NIQE scores indicate better performance.

\begin{figure*}[t]
  \centering
  \includegraphics[width=\textwidth]{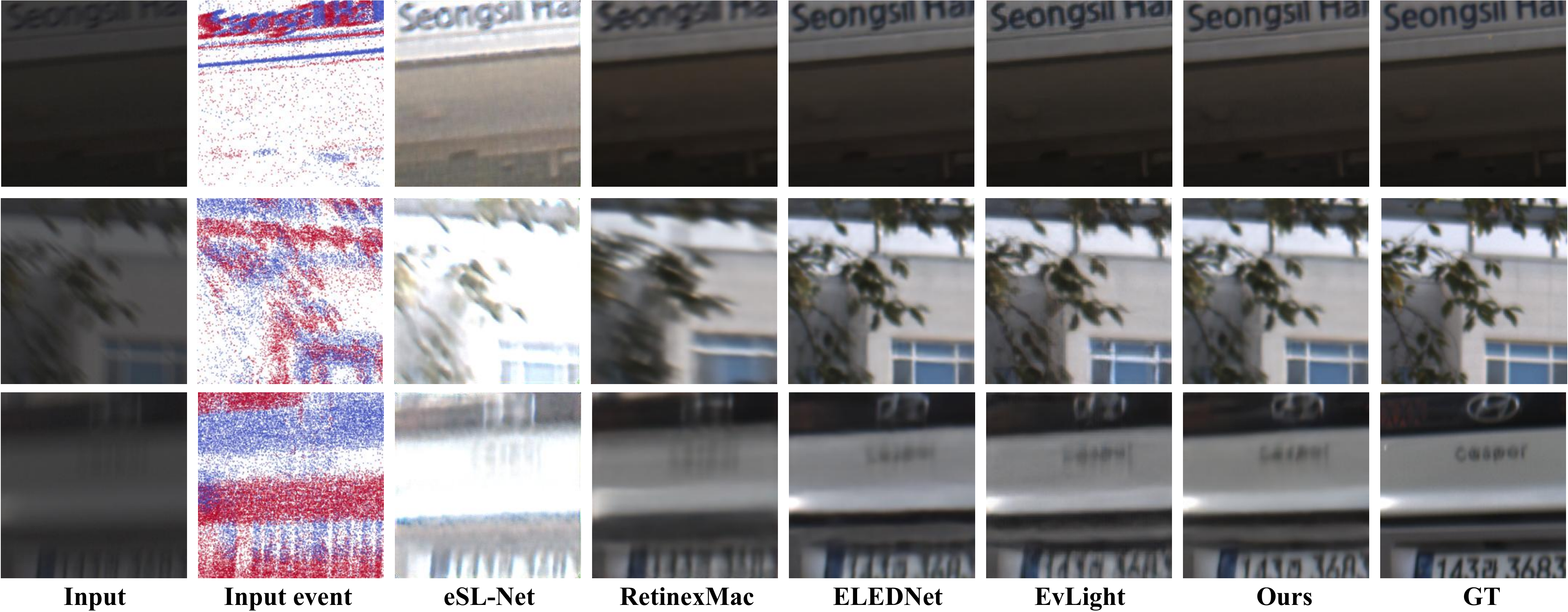} 
  \caption{Visual comparisons  with state-of-the-art methods on the RELED dataset \cite{kim2024towards}.}
  \label{fig:reled}
\end{figure*}

\begin{figure}[!t] 
  \centering
  \includegraphics[width=\columnwidth]{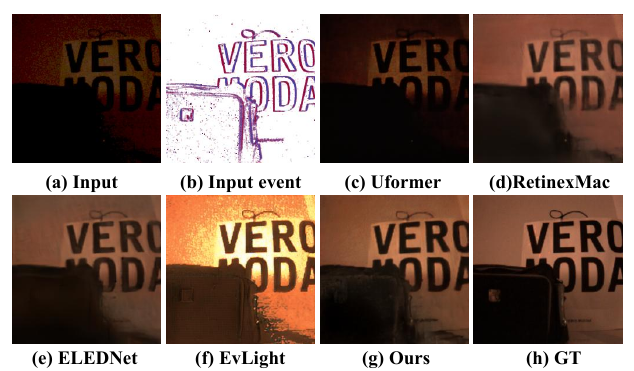}
  \caption{Cross-dataset generalization results on the LIE \cite{jiang2023event} dataset using the model trained on SDE-out.
}
  \label{fig:lie}
\end{figure}

\begin{table}[t]
\renewcommand{\arraystretch}{1.3}
\centering
\fontsize{8.5}{11}\selectfont
\caption{Cross-dataset generalization on LIE (trained on SDE-out). The best 
and second-best results are in bold and underlined.}
\resizebox{\columnwidth}{!}{
\begin{tabular}{
>{\centering\arraybackslash}p{1cm}
>{\centering\arraybackslash}p{1.8cm}
>{\centering\arraybackslash}p{2cm}
>{\centering\arraybackslash}p{2cm}
>{\centering\arraybackslash}p{1.8cm}
>{\centering\arraybackslash}p{1.5cm}
}
\Xhline{1.5pt}
\textbf{Method} & \textbf{Uformer}  \cite{wang2022uformer} & \textbf{RetinexMac} \cite{10810469}& \textbf{ELEDNet} \cite{kim2024towards}& \textbf{EvLight} \cite{liang2024towards}& \textbf{Ours} \\
\hline
\textbf{PSNR$\uparrow$} & 17.20 & \underline{20.68} & 20.55 & 11.48 & \textbf{20.75} \\
\textbf{SSIM$\uparrow$} & 0.388 & \underline{0.638} & 0.637 & 0.285 & \textbf{0.645} \\
\Xhline{1.5pt}
\end{tabular}}
\label{tab:lie}
\end{table}

\subsubsection{Implementation Details}

All experiments were conducted with the Adam optimizer \cite{kingma2014adam}. Learning rates were set to \(1 e{-4}\) for \textsc{SDE} datasets, \(1.5 e{-4}\) for \textsc{SDSD} datasets, and \(1 e{-4}\) for \textsc{RELED} datasets. The framework was trained on an NVIDIA RTX 3090 GPU for 80 training cycles with a batch size of eight. For the training set, we employed data-augmentation techniques consisting of random \(256 \times 256\) crops, horizontal flips, and rotations of \(90^{\circ}\), \(180^{\circ}\), and \(270^{\circ}\). For the test set, we applied only a center crop of \(256 \times 256\).

\subsubsection{State-of-the-Art Methods for Comparisons}

We compared our EvRWKV with ten methods, including five methods that only use RGB images as input(SNR-Net \cite{xu2022snr}, Uformer \cite{wang2022uformer}, LLFlow-L-SKF \cite{wu2023learning}, RetinexMac \cite{10810469}, Retinexformer \cite{cai2023retinexformer}), five methods that use a RGB image and paired events as inputs(ELIE \cite{jiang2023event}, eSL-Net \cite{wang2020event}, LLVE-SEG \cite{liu2023low}, ELEDNet \cite{kim2024towards}, EvLight \cite{liang2024towards}). Since the source code of LLVE-SEG \cite{liu2023low} is not publicly available, we implemented it according to the description in its paper. For the SNR-Net \cite{xu2022snr}, Uformer \cite{wang2022uformer}, LLFlow-L-SKF \cite{wu2023learning}, RetinexMac \cite{10810469}, Retinexformer \cite{cai2023retinexformer}, ELIE \cite{jiang2023event}, eSL-Net \cite{wang2020event}, ELEDNet \cite{kim2024towards} and EvLight \cite{liang2024towards} methonds, we used the codes released by the authors to output their results.

\subsection{Comparison and Evaluation}

We evaluate the proposed EvRWKV framework on three challenging datasets, namely SDE, SDSD, and RELED, covering diverse LLIE scenarios. SDE and RELED contain real-world event-image pairs captured under indoor and outdoor 
low-light conditions as well as motion blur scenarios, while SDSD consists of synthetic events generated from dynamic video sequences. We use PSNR, SSIM, LPIPS, and NIQE as quantitative metrics to assess both pixel-level accuracy and perceptual quality, and provide detailed qualitative analysis to validate the effectiveness of our approach.

\subsubsection{Evaluation on the SDE Dataset}

For LLIE on the SDE dataset, the ability to simultaneously suppress noise and preserve fine structural details is the key evaluation criterion. We first qualitatively compare our EvRWKV method with several representative approaches including eSL-Net \cite{wang2020event}, ELEDNet \cite{kim2024towards}, EvLight \cite{liang2024towards}, and RetinexMac \cite{10810469}. As shown in Fig.~\ref{fig:sde}, RetinexMac and ELEDNet reduce noise to some extent but tend to oversmooth textures, resulting in loss of important details in dark regions. As shown in Fig.~\ref{fig:sde}, EvLight maintains more structural information but suffers from residual noise and blur artifacts. eSL-Net, with a relatively small model size, provides limited enhancement capabilities and often underperforms in challenging scenes. In contrast, our EvRWKV method achieves a better balance between noise suppression and texture preservation, producing cleaner, brighter images with rich details and minimal artifacts. This demonstrates the effectiveness of our dual-domain fusion and adaptive gating mechanisms in handling complex low-light scenarios.

In terms of quantitative evaluation, Table~\ref{tab:1} presents comprehensive metrics for different methods on the SDE dataset. Our method achieves the highest PSNR of 23.09 and SSIM of 0.770 for SDE-in, along with the best LPIPS of 0.094 and NIQE of 8.534, demonstrating superior performance in both pixel-level reconstruction and perceptual quality. For SDE-out, our method obtains 23.43 PSNR, 0.768 SSIM, 0.091 LPIPS, and 8.926 NIQE, consistently outperforming all compared methods including SNR-Net \cite{xu2022snr}, Uformer \cite{wang2022uformer}, LLFlow \cite{wu2023learning}, Retinexformer \cite{cai2023retinexformer}, ELIE \cite{jiang2023event}, 
EvLight \cite{liang2024towards}, and ELEDNet \cite{kim2024towards}. The significant improvements across both traditional metrics and perceptual metrics validate the superiority of our EvRWKV framework in LLIE.

\subsubsection{Evaluation on the SDSD Dataset}  
We further qualitatively compare the performance of different methods on the SDSD dataset for LLIE under challenging lighting and motion conditions. In Fig.~\ref{fig:sdsd}, we observe that most methods improve the visibility and brightness of degraded images. However,  RetinexMac \cite{10810469}, ELEDNet \cite{kim2024towards}, and EvLight \cite{liang2024towards} exhibit limitations: RetinexMac and ELEDNet tend to over-smooth textures and lose fine details, While EvLight, although preserving more structural information, causes noticeable color distortion in some regions. As shown in Fig.~\ref{fig:sdsd}, eSL-Net \cite{wang2020event} shows relatively limited enhancement capability due to its smaller model size. In comparison, our EvRWKV method delivers superior noise suppression and detail preservation, effectively enhancing both static and dynamic scene elements without introducing artifacts. The enhanced images show clearer textures, better contrast, and natural brightness, demonstrating the robustness of our dual-domain cross-modal fusion strategy.

Table~\ref{tab:2} presents comprehensive quantitative evaluation on the SDSD dataset. Our method achieves the best performance across multiple metrics in both SDSD-in and SDSD-out scenarios. Specifically, for SDSD-in, we obtain PSNR of 28.96, SSIM of 0.920, LPIPS of 0.032, and NIQE of 6.451, while for SDSD-out, we achieve 27.93 PSNR, 0.839 SSIM, 0.058 LPIPS, and 4.110 NIQE. These results consistently outperform all compared methods, demonstrating superior performance in both pixel-level accuracy and perceptual quality. The improvements in LPIPS and NIQE further validate that our enhanced images better align with human visual perception and exhibit more natural image characteristics. Overall, our approach demonstrates both qualitative and quantitative advantages over existing 
state-of-the-art techniques on the SDSD dataset.

\subsubsection{Evaluation on the RELED Dataset}  
To comprehensively evaluate the enhancement performance of our EvRWKV method on the RELED dataset, we compare it qualitatively with several representative approaches including eSL-Net \cite{wang2020event}, ELEDNet \cite{kim2024towards}, EvLight \cite{liang2024towards}, and RetinexMac \cite{10810469}. As shown in Fig.~\ref{fig:reled}, the RELED dataset contains extremely low-light images with severe motion blur and noise. As shown in Fig.~\ref{fig:reled}, eSL-Net often causes overexposure while attempting to enhance brightness. RetinexMac boosts overall brightness but fails to eliminate noticeable blur. ELEDNet and EvLight aim to balance illumination, but this often leads to the loss of finer details. In contrast, our EvRWKV method effectively leverages event-based motion cues and image textures to suppress noise, reduce blur, and restore fine details, producing visually clearer, sharper, and more natural images without color artifacts or overexposure.

Table~\ref{tab:2} reports comprehensive metrics for different methods on the RELED dataset. Our EvRWKV achieves the best quantitative results with PSNR of 32.18, SSIM of 0.911, LPIPS of 0.044, and NIQE of 9.207, significantly outperforming other methods including eSL-Net \cite{wang2020event}, ELEDNet \cite{kim2024towards}, EvLight \cite{liang2024towards}, and RetinexMac \cite{10810469}. These comprehensive quantitative gains across both traditional and perceptual metrics demonstrate the robustness and generalization capability of our dual-domain cross-modal fusion framework in handling challenging low-light and motion-degraded scenarios. Overall, our method achieves excellent qualitative and quantitative results across all three real-world LLIE benchmarks.

\begin{table}[!t]
\renewcommand{\arraystretch}{1.2}
\centering
\fontsize{8.5}{11}\selectfont
\caption{Ablation study of key components on the SDSD-in dataset.}
\resizebox{\columnwidth}{!}{
\begin{tabular}{
>{\centering\arraybackslash}p{1.4cm}
>{\centering\arraybackslash}p{2cm}
>{\centering\arraybackslash}p{2cm}
>{\centering\arraybackslash}p{1.2cm}
>{\centering\arraybackslash}p{1.2cm}
}
\Xhline{1.5pt}
\textbf{EISFE} & \textbf{SpatialMix} & \textbf{ChannelMix} & \textbf{PSNR$\uparrow$} & \textbf{SSIM$\uparrow$} \\
\hline
\ding{55} & \ding{51} & \ding{51} & 27.35 & 0.894 \\
\ding{51} & \ding{55} & \ding{51} & 26.88 & 0.798 \\
\ding{51} & \ding{51} & \ding{55} & 26.60 & 0.896 \\
\ding{51} & \ding{51} & \ding{51} & \textbf{28.96} & \textbf{0.920} \\
\Xhline{1.5pt}
\end{tabular}}
\label{tab:3}
\end{table}

\begin{table}[!t]
\renewcommand{\arraystretch}{1.3}
\centering
\fontsize{8.5}{11}\selectfont
\caption{Ablation study of loss functions on the SDSD-in dataset.}
\resizebox{\columnwidth}{!}{
\begin{tabular}{
>{\centering\arraybackslash}p{1.5cm}
>{\centering\arraybackslash}p{1.5cm}
>{\centering\arraybackslash}p{1.5cm}
>{\centering\arraybackslash}p{1.5cm}
>{\centering\arraybackslash}p{2cm}
>{\centering\arraybackslash}p{2cm}
}
\Xhline{1.5pt}
\textbf{$\lambda_r$} & \textbf{$\lambda_p$} & \textbf{$\lambda_s$} & \textbf{$\lambda_m$} & \textbf{PSNR$\uparrow$} & \textbf{SSIM$\uparrow$} \\
\hline
0 & 0.8 & 1 & 0.5 & 26.84 & 0.896 \\
1 & 0 & 1 & 0.5 & 26.18 & 0.885 \\
1 & 0.8 & 0 & 0.5 & 26.06 & 0.876 \\
1 & 0.8 & 1 & 0 & 27.84 & 0.905 \\
1 & 0.8 & 1.5 & 0.5 & 27.98 & 0.908 \\
1 & 0.8 & 1.5 & 1 & 26.12 & 0.891\\
1 & 1 & 1 & 1 & 26.73 & 0.891 \\
1 & 0.8 & 1 & 0.5 & \textbf{28.96} & \textbf{0.920} \\
\Xhline{1.5pt}
\end{tabular}}
\label{tab:4}
\end{table}

\begin{table}[!t]
\renewcommand{\arraystretch}{1.3}
\centering
\fontsize{8.5}{11}\selectfont
\caption{Ablation study of voxel bin size on the RELED dataset.}
\resizebox{0.7\columnwidth}{!}{
\begin{tabular}{
>{\centering\arraybackslash}p{4.5cm}
>{\centering\arraybackslash}p{2cm}
>{\centering\arraybackslash}p{2cm}
}
\Xhline{1.5pt}
\textbf{Voxel Bin Size} & \textbf{PSNR$\uparrow$} & \textbf{SSIM$\uparrow$} \\
\hline
$B = 16$ & 30.42 & 0.879 \\
$B = 64$ & 30.57 & 0.881 \\
$B = 32$ (Ours) & \textbf{32.18} & \textbf{0.911} \\
\Xhline{1.5pt}
\end{tabular}}
\label{tab:voxel}
\end{table}

\begin{table}[!t]
\renewcommand{\arraystretch}{1.3}
\centering
\fontsize{8.5}{11}\selectfont
\caption{Ablation study of Cross-RWKV depth on the SDSD-in dataset.}
\resizebox{\columnwidth}{!}{
\begin{tabular}{
>{\centering\arraybackslash}p{1cm}
>{\centering\arraybackslash}p{3cm}
>{\centering\arraybackslash}p{1.5cm}
>{\centering\arraybackslash}p{1.5cm}
>{\centering\arraybackslash}p{1.8cm}
}
\Xhline{1.5pt}
\textbf{Levels} & \textbf{Block Configuration} & \textbf{PSNR$\uparrow$} & \textbf{SSIM$\uparrow$} & \textbf{Params (M)} \\
\hline
3 & [1, 2, 2] & 25.63 & 0.891 & 4.3 \\
3 & [2, 4, 4] & 26.04 & 0.896 & 6.6 \\
5 & [1, 2, 2, 4, 4] & \textbf{29.02} & 0.911 & 82.2 \\
5 & [2, 4, 4, 8, 8] & 27.60 & 0.898 & 147 \\
4 & [2, 4, 4, 8] & 28.18 & 0.901 & 33.1 \\
4 (Ours) & [1, 2, 2, 6] & 28.96 & \textbf{0.920} & 24.2 \\
\Xhline{1.5pt}
\end{tabular}}
\label{tab:level}
\end{table}

\subsubsection{Generalization Comparison }
\label{sec:generalization}
To evaluate the generalization capability and analyze the limitations of our method, we conduct cross-dataset experiments using the model trained on SDE-out and directly testing it on the LIE dataset \cite{jiang2023event} without any fine-tuning. Table~\ref{tab:lie} shows that our method achieves the best performance with PSNR of 20.75 and SSIM of 0.645. As shown in Fig.~\ref{fig:lie}, our method effectively recovers details in low-light images, while simultaneously preventing overexposure, as seen in (e), and color distortion, as seen in (d). These results demonstrate the strong cross-dataset generalization capability and robustness of EvRWKV.

\begin{figure}[!t]
  \centering
  \includegraphics[width=\columnwidth]{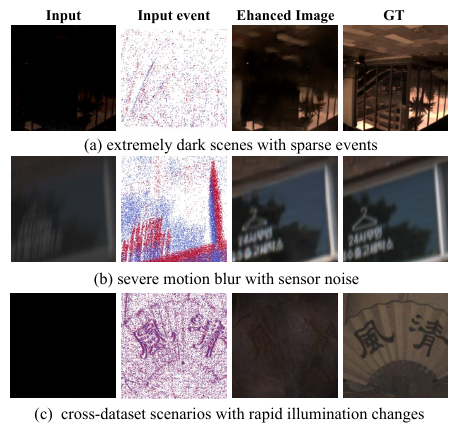}
  \caption{Representative failure cases in challenging scenarios.}
  \label{fig:failure}
\end{figure}

\subsection{Ablation Study}
We conduct comprehensive ablation studies to evaluate the contribution of each key component and hyper-parameter in our EvRWKV framework. We assess the effectiveness of the EISFE module, the SpatialMix and ChannelMix mechanisms 
within the Cross-RWKV Module, the role of each loss term, as well as the influence of critical hyper-parameters including Cross-RWKV depth and voxel bin size. For each study, we isolate the impact by removing or modifying a single component while keeping the rest unchanged. The results clearly demonstrate the necessity and complementarity of these components in achieving robust and high-quality LLIE. As shown in Fig.~\ref{fig:wo}, the removal of EISFE (b) results in color distortion. Without SpatialMix (c), visible noise and blurred textures appear, while omitting ChannelMix (d) leads to flat tonal rendering. The full model (e) demonstrates the best visual clarity and color fidelity, especially in complex shadow regions.

\subsubsection{Impact of EISFE Module}
To assess the contribution of frequency-aware fusion, we remove the EISFE module and directly decode the features after the Cross-RWKV backbone. As shown in Table~\ref{tab:3}, the absence of EISFE leads to a 1.61 dB drop in PSNR and 0.026 reduction in SSIM. This indicates the importance of frequency-domain filtering and spatial deformable convolution in denoising and structure restoration. Visually, the model without EISFE shows residual noise and blurred edges.

\subsubsection{Impact of SpatialMix and ChannelMix}
\begin{figure}[!t] 
  \centering
  \includegraphics[width=\columnwidth]{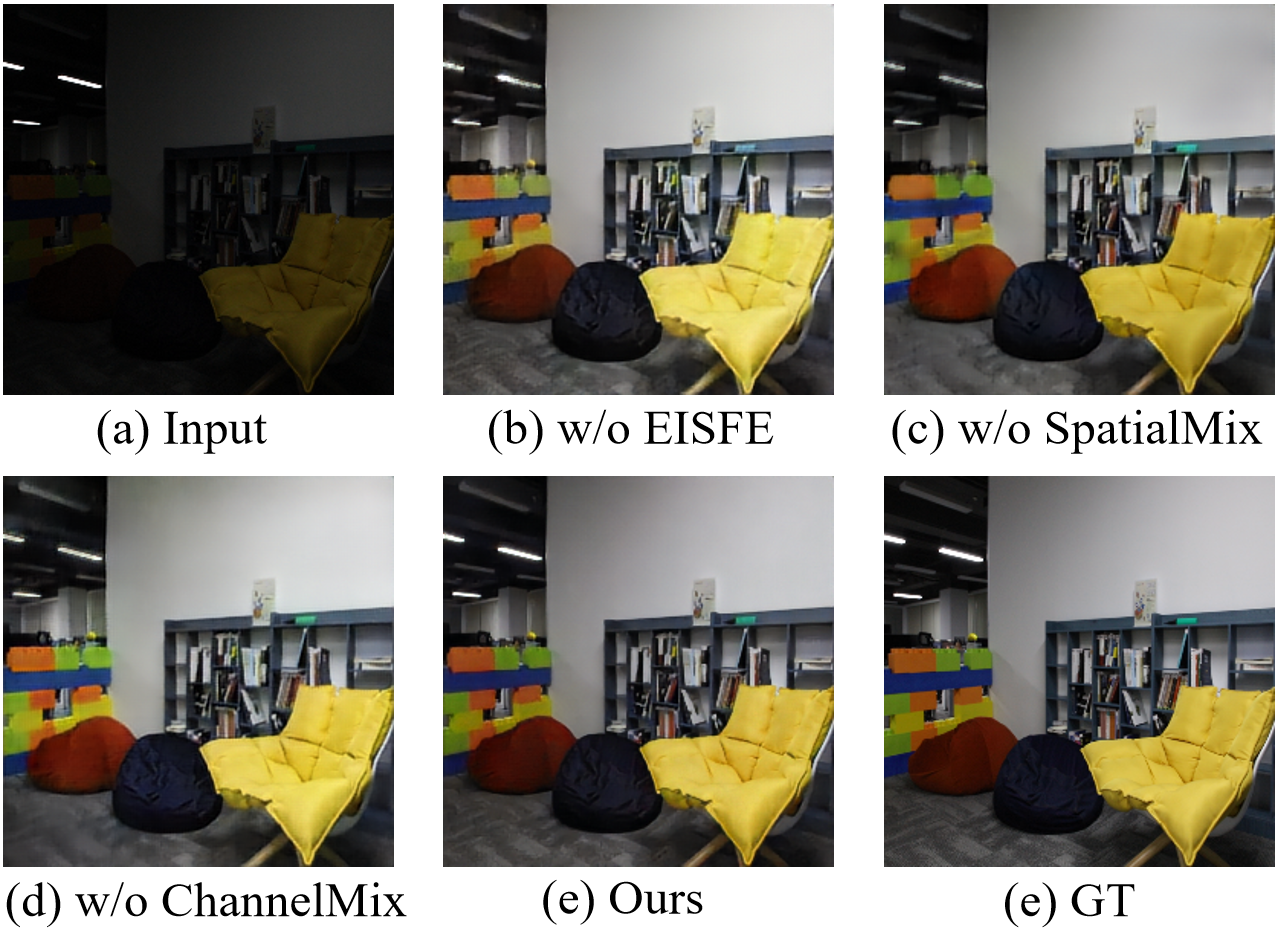} 
  \caption{Visualization of ablation results.
}
  \label{fig:wo}
\end{figure}

To verify the effectiveness of the spatial and channel interaction designs in Cross-RWKV, we conduct ablation studies by individually removing each module. As shown in Table III, removing the SpatialMix results in a 2.08 dB PSNR drop and 0.122 SSIM decline, highlighting its critical role in capturing long-range dependencies and facilitating event-image alignment. On the other hand, removing ChannelMix leads to a 2.36 dB decrease in PSNR and a 0.024 reduction in SSIM, indicating its importance in refining semantic consistency across feature channels. These results confirm that both modules contribute distinctly and significantly to performance. As illustrated in Fig.~\ref{fig:wo}, ChannelMix aids in preserving local contrast, while SpatialMix enhances spatial structure and reduces texture inconsistency.

\subsubsection{Impact of Loss Functions}  
To validate our loss configuration, we conducted ablation studies by both removing individual loss terms and adjusting their weights, as detailed in Table~\ref{tab:4}. 
Removing either the SSIM loss $L_s$ or the perceptual loss $L_p$ causes a significant performance drop to 26.06 dB and 26.18 dB respectively, highlighting their crucial role in preserving structural and perceptual quality. 
While other configurations with different weightings show competitive results, our final empirically determined weights  yield the optimal performance of 28.96 dB PSNR and 0.920 SSIM. This confirms that both the selection and weighting of the loss terms are critical for optimal results. 

\subsubsection{Impact of Voxel Bin Size}
The voxel bin size $B$ determines the temporal resolution of event representation. We evaluate three settings on the RELED dataset. As shown in Table~\ref{tab:voxel}, $B = 16$ yields a PSNR of 30.42 dB, while $B = 64$ achieves 30.57 dB. Our default setting $B = 32$ achieves the best performance with 32.18 dB PSNR and 0.911 SSIM. This indicates that $B = 32$ provides an optimal balance: smaller bins suffer from excessive sparsity, while larger bins lose temporal precision.

\subsubsection{Impact of Cross-RWKV Depth}
The number of stacked Cross-RWKV blocks affects the model's capacity to capture long-range dependencies. As shown in Table~\ref{tab:level}, using 3 levels results in 25.63 dB PSNR with only 4.3M parameters, indicating insufficient capacity. Increasing to 5 levels achieves 29.02 dB but requires 82.2M parameters. Our 4-level configuration achieves 28.96 dB PSNR and 0.920 SSIM with 24.2M parameters, providing the best balance between performance and computational cost.


\begin{figure*}[!t]
  \centering
  \includegraphics[width=\textwidth]{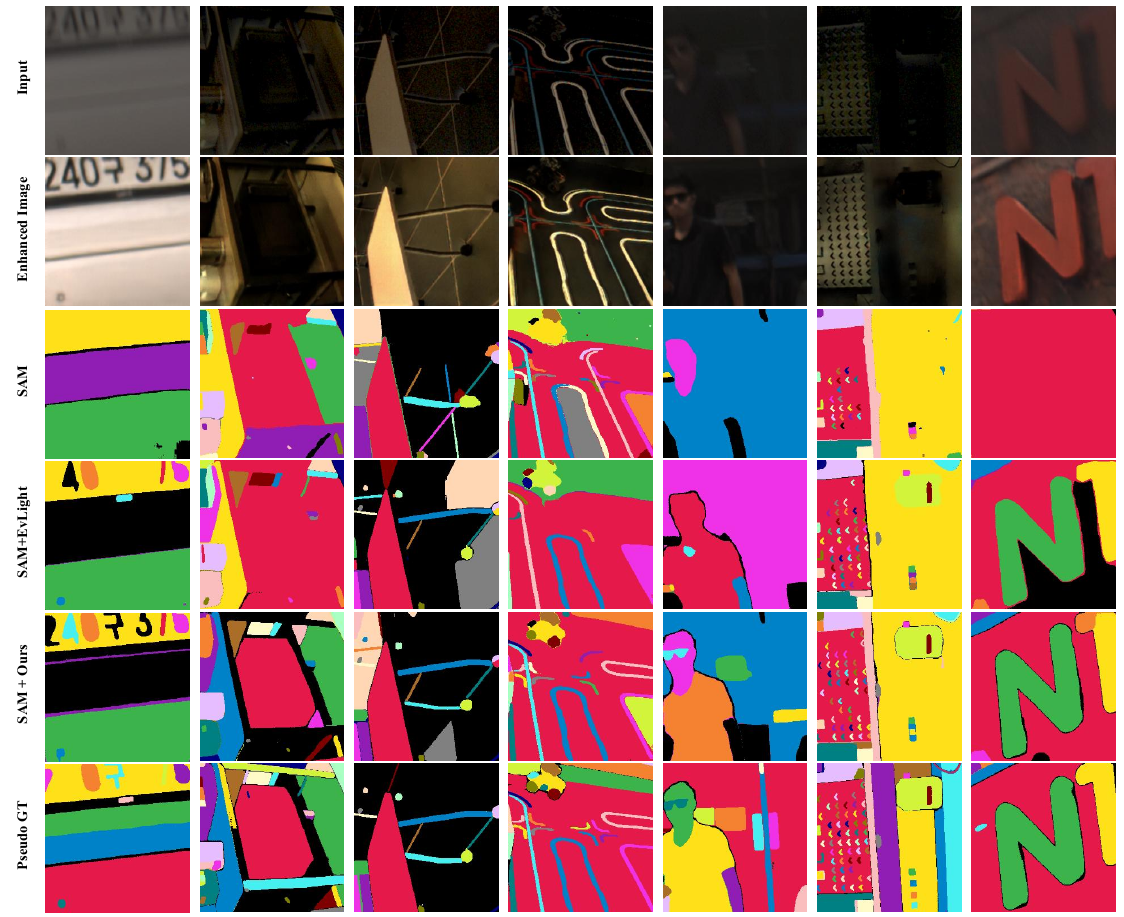} 
  \caption{Qualitative comparison of semantic segmentation results for low-light scenes. The baseline method (i.e., SAM \cite{kirillov2023segment}) uses the raw low-light image as input, SAM+EvLight uses the EvLight-enhanced image as input, while SAM+Ours uses our enhanced image as input. GT shows the segmentation on GT images.}
  \label{fig:seg}
\end{figure*}

\begin{table}[!t]
\renewcommand{\arraystretch}{1.3}
\centering
\fontsize{8.5}{11}\selectfont
\caption{Quantitative results of semantic segmentation on our SDE-in and RELED dataset.}
\resizebox{\columnwidth}{!}{
\begin{tabular}{
>{\centering\arraybackslash}p{2.5cm}
>{\centering\arraybackslash}p{1.2cm}
>{\centering\arraybackslash}p{1.2cm}
>{\centering\arraybackslash}p{1.2cm}
>{\centering\arraybackslash}p{1.2cm}
>{\centering\arraybackslash}p{1.2cm}
>{\centering\arraybackslash}p{1.2cm}
}
\specialrule{1.5pt}{0pt}{0pt}
\multirow{2}{*}{\textbf{Method}} & \multicolumn{3}{c}{\textbf{SDE-in}} & \multicolumn{3}{c}{\textbf{RELED}} \\
\cline{2-7}
& \textbf{aAcc$\uparrow$} & \textbf{mIoU$\uparrow$} & \textbf{mAcc$\uparrow$} & \textbf{aAcc$\uparrow$} & \textbf{mIoU$\uparrow$} & \textbf{mAcc$\uparrow$} \\
\midrule
SAM \cite{kirillov2023segment} & 60.46 & 30.84 & 40.81 & 68.64 & 20.13 & 29.13 \\
SAM \cite{kirillov2023segment}+EvLight & 66.38 & 38.16 & 49.52 & 78.26 & 37.27 & 46.70 \\
SAM \cite{kirillov2023segment}+Ours & \textbf{69.27} & \textbf{41.77} & \textbf{52.55} & \textbf{80.90} & \textbf{39.37} & \textbf{48.25} \\
\midrule
Improve (\%) & \textcolor{blue}{+4.35} & \textcolor{blue}{+9.46} & \textcolor{blue}{+6.12} & \textcolor{blue}{+3.37} & \textcolor{blue}{+5.63} & \textcolor{blue}{+3.32} \\
\specialrule{1.5pt}{0pt}{0pt}
\end{tabular}}
\label{tab:downstream}
\end{table}

\subsection{Downstream Applications}
\label{sec:downstream}

While event cameras offer unique advantages for low-light perception, leveraging their complementary information to enhance frame quality enables the direct application of well-established frame-based vision algorithms without requiring specialized event-based processing. To demonstrate the practical utility of our enhancement method, we conduct semantic segmentation experiments using the off-the-shelf ViT-H SAM model \cite{kirillov2023segment} on the SDE-in and RELED datasets. We compare segmentation performance across four types of inputs: raw low-light images, EvLight-enhanced images \cite{liang2024towards}, our enhanced images, and ground-truth (GT). Since pixel-level segmentation annotations are not available, we use SAM segmentation results on GT as pseudo GT, which provides a reliable reference given SAM's optimal performance on high-quality, well-exposed images. For quantitative evaluation, we report average pixel accuracy (aAcc), mean Intersection over Union (mIoU), and mean class accuracy (mAcc) in Table~\ref{tab:downstream}. As shown in Fig.~\ref{fig:seg}, our enhancement better preserves discriminative features in challenging regions with fine-grained structures compared to both raw images and EvLight, producing segmentation masks with greater consistency to the GT. These results confirm that our EvRWKV effectively recovers semantic content essential for practical applications such as autonomous driving and robotic navigation in low-light environments.

\subsection{Limitations}
\label{label:limitations}
Despite the strong performance, EvRWKV encounters challenges in certain 
extreme scenarios, as illustrated in Fig.~\ref{fig:failure}. First, 
in extremely dark scenes where event activity becomes highly sparse due to 
minimal brightness changes, our method struggles to extract sufficient motion 
cues for effective cross-modal fusion. As shown in 
Fig.~\ref{fig:failure}(a), the scarcity of events in such conditions 
prevents our Cross-RWKV Module from establishing reliable correspondences 
with image features, occasionally leading to over-smoothing or color artifacts. 
Second, when RGB frames suffer from severe motion blur combined with sensor 
noise, both modalities are simultaneously degraded. As illustrated in Fig.~\ref{fig:failure}(b), since event voxels aggregate multiple 
temporal instances into a single frame representation, critical motion 
information is lost during this temporal compression, limiting our method's 
ability to recover sharp details from blurred inputs. Third, in cross-dataset 
scenarios with rapid illumination changes, as shown in Fig.~\ref{fig:failure}(c), sudden lighting transitions  trigger bursts of events that do not correspond to actual 
scene structure, interfering with our spatial-temporal alignment and resulting 
in degraded outputs. These observations motivate future work on more robust event processing and domain adaptation techniques to overcome these extreme scenarios.

\section{Conclusion}
\label{label:conclusion}
In this paper, we proposed EvRWKV, an effective framework for LLIE that establishes continuous cross-modal interaction between event streams and low-light images. The Cross-RWKV Module, leveraging a recurrent structure, maintains feature consistency from low-level textures to high-level semantics, directly addressing the information loss and representational bottleneck issues of prior paradigms. This is complemented by the EISFE module, which performs robust dual-domain fusion by jointly handling frequency-domain noise and spatial-domain alignment
Extensive experiments validate our approach. EvRWKV significantly outperforms state-of-the-art methods across both quantitative metrics and visual comparisons, effectively restoring fine details while suppressing severe artifacts.
In future work, we identify two specific and promising directions. A primary avenue is the extension of EvRWKV to real-time video enhancement, where the recurrent structure and efficiency of our framework provide a strong foundation for maintaining temporal consistency. A second, more challenging direction is to improve robustness under extreme low-light conditions where event streams become exceedingly sparse, which may require novel strategies such as integrating generative priors to compensate for the data scarcity.

\bibliographystyle{IEEEtran}
\bibliography{reference}

\vspace{-40pt}
\begin{IEEEbiography}[{\includegraphics[width=1in,height=1.25in,clip,keepaspectratio]{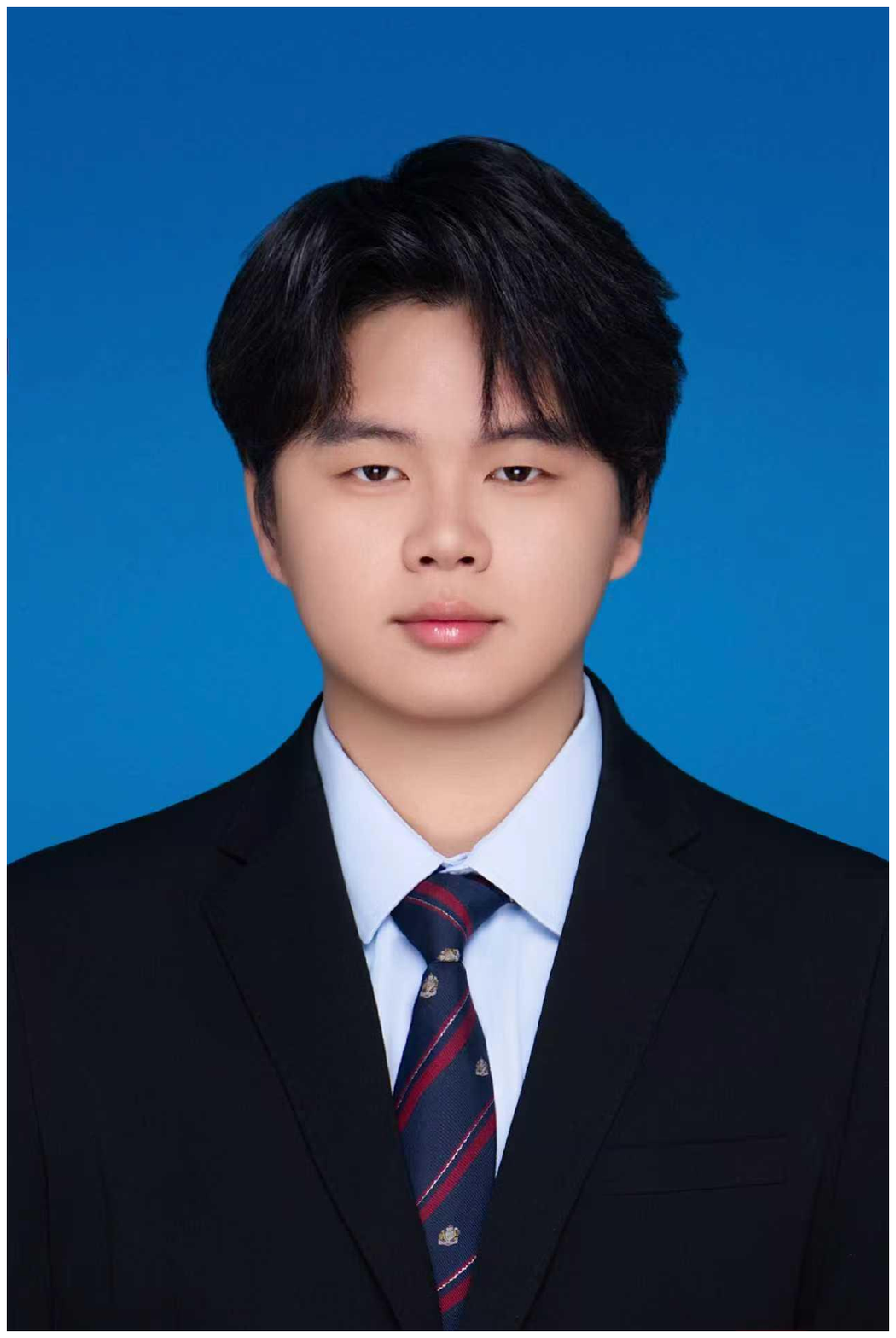}}]{Wenjie Cai}
is currently pursuing the B.E. degree in Artificial Intelligence with the School of Artificial Intelligence, Anhui University, Hefei, China. He serves as a Research Intern at the Anhui Provincial International Joint Research Center for Advanced Technology in Medical Imaging.
His research interests include image and video processing, computer vision and deep learning.
\end{IEEEbiography}

\begin{IEEEbiography}[{\includegraphics[width=1in,height=1.25in,clip,keepaspectratio]{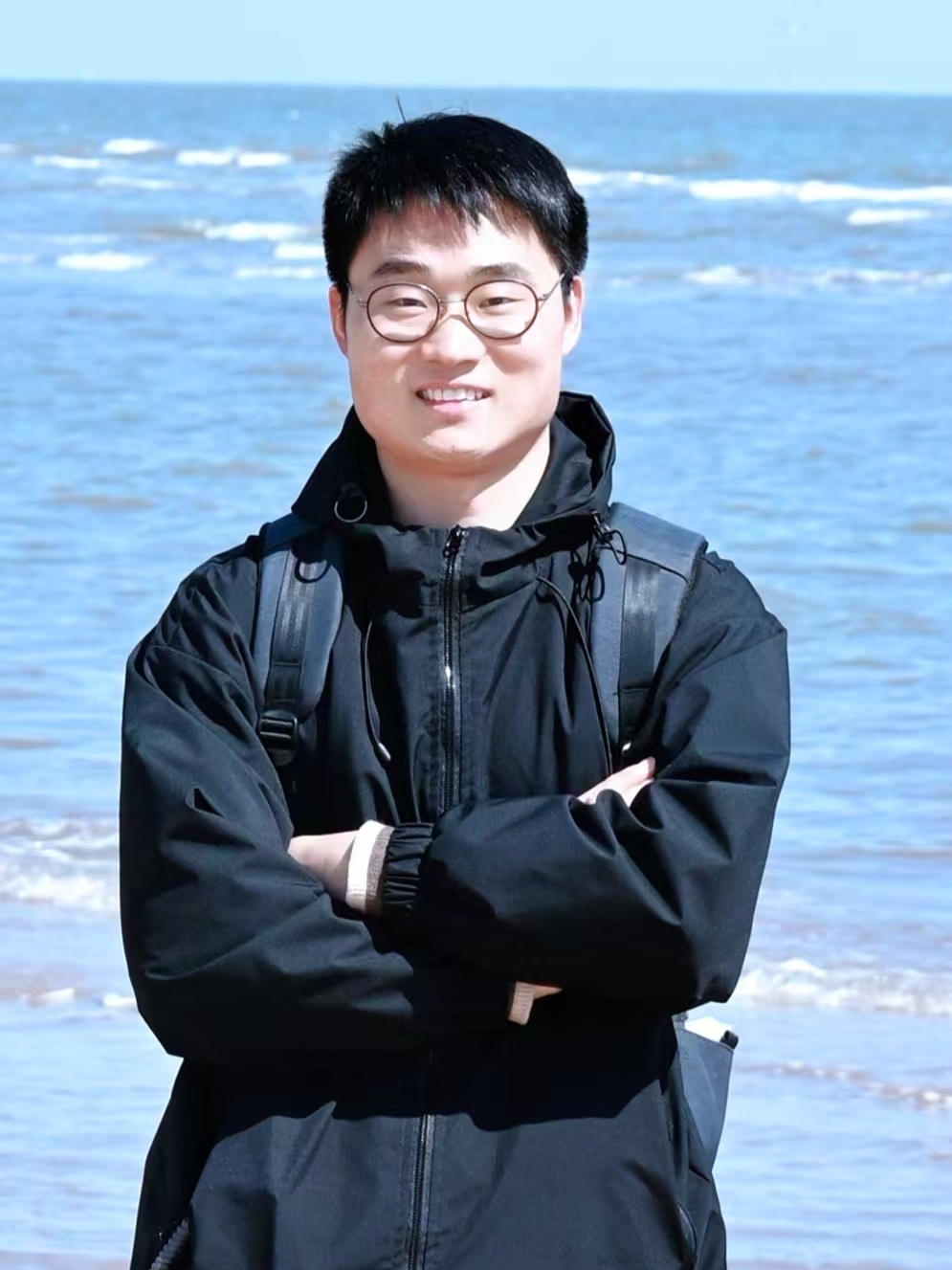}}]{Qingguo Meng}
received his B.Eng. degree in computer science and technology from Henan Polytechnic University, Jiaozuo, China, in 2022. He is currently working on his Ph.D. at the School of Artificial Intelligence, Anhui University in Hefei, China. His research directions are object tracking, low-light image enhancement, medical imaging, and biometrics.
\end{IEEEbiography}

\begin{IEEEbiography}[{\includegraphics[width=1in,height=1.25in,clip,keepaspectratio]{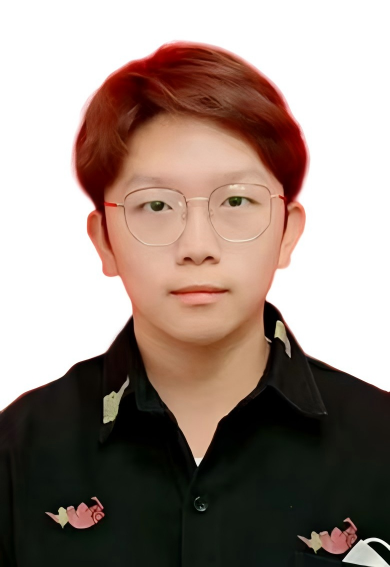}}]{Zhenyu Wang}
is currently pursuing the B.E. degree in Artificial Intelligence with the School of Artificial Intelligence, Anhui University, Hefei, China. His research interests include image processing and computer vision.
\end{IEEEbiography}
\begin{IEEEbiography}[{\includegraphics[width=1in,height=1.25in,clip,keepaspectratio]{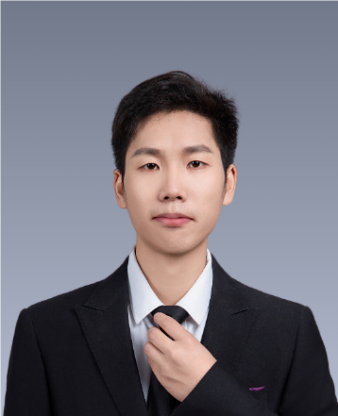}}]{Xingbo Dong} (Member, IEEE) received the B.S. degree from Huazhong Agriculture University, Wuhan, China, in 2014, and the Ph.D. degree from the Faculty of Information Technology, Monash University, Melbourne, VIC, Australia, in 2021.
He held a post-doctoral position with Yonsei University, Seoul, South Korea, in 2022. He is currently a Lecturer with Anhui University, Hefei, China. His research interests include biometrics, medical imaging, and image processing.
\end{IEEEbiography}

\begin{IEEEbiography}[{\includegraphics[width=1in,height=1.25in,clip,keepaspectratio]{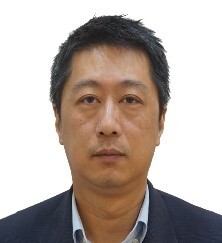}}]{Zhe Jin}
(Member, IEEE) obtained a Ph.D. in Engineering from Universiti Tunku Abdul Rahman Malaysia (UTAR). He is a Professor at the School of Artificial Intelligence, Anhui University, China. His research interests include Biometrics, Pattern Recognition, Computer Vision, and Multimedia Security. He has published over 70 refereed journals and conference articles, including IEEE Trans. IFS, SMC-S, DSC, PR. He was awarded the Marie Skłodowska-Curie Research Exchange Fellowship. He visited the University of Salzburg, Austria, and the University of Sassari, Italy, respectively, as a visiting scholar under the EU Project IDENTITY 690907.
\end{IEEEbiography}
\end{document}